\documentclass[11pt]{article}
\usepackage{a4p,cite}
\newcommand{\rb}{\mbox{$R_{\rm b}$}}
\newcommand{\rc}{\mbox{$R_{\rm c}$}}
\newcommand{\zb}{\mbox{$\rm Z^0$}}

\newcommand{\ccbar}{\mbox{$\rm c\overline{c}$}}
\newcommand{\bbbar}{\mbox{$\rm b\overline{b}$}}
\newcommand{\qqbar}{\mbox{$\rm q\overline{q}$}}
\newcommand{\bqbar}{\mbox{$\rm\overline{b}$}}

\newcommand{\eudst}{\mbox{$\epsilon_{\rm T}^{\rm uds}$}}
\newcommand{\ect}{\mbox{$\epsilon_{\rm T}^{\rm c}$}}
\newcommand{\ebt}{\mbox{$\epsilon_{\rm T}^{\rm b}$}}
\newcommand{\eudsm}{\mbox{$\epsilon_{\rm M}^{\rm uds}$}}
\newcommand{\ecm}{\mbox{$\epsilon_{\rm M}^{\rm c}$}}
\newcommand{\ebm}{\mbox{$\epsilon_{\rm M}^{\rm b}$}}
\newcommand{\tmpure}{\mbox{$\Pi_{\rm TM}$}}
\newcommand{\mean}[1]{\langle{#1}\rangle}

\newcommand{\bplus}{\mbox{$\rm B^+$}}
\newcommand{\bminus}{\mbox{$\rm B^-$}}
\newcommand{\bzero}{\mbox{$\rm B^0$}}
\newcommand{\bzerobar}{\mbox{$\rm\bar{B}^0$}}
\newcommand{\bx}{\mbox{$\rm B_1$}}
\newcommand{\by}{\mbox{$\rm B_2$}}
\newcommand{\bs}{\mbox{$\rm B_s$}}
\newcommand{\bsbar}{\mbox{$\rm \bar{B}_s$}}
\newcommand{\bbary}{\mbox{$\rm \Lambda_b$}}
\newcommand{\bbarybar}{\mbox{$\rm \bar{\Lambda}_b$}}
\newcommand{\taubp}{\mbox{$\rm\tau_{B^+}$}}
\newcommand{\taubm}{\mbox{$\rm\tau_{B^-}$}}
\newcommand{\taubz}{\mbox{$\rm\tau_{B^0}$}}
\newcommand{\taubr}{\mbox{$\rm\tau_{B^+}/\tau_{B^0}$}}
\newcommand{\taubg}{\mbox{$\rm\tau_{\rm bg}$}}
\newcommand{\taub}{\mbox{$\rm\tau_{b}$}}
\newcommand{\taubbar}{\mbox{$\rm\tau_{\bar{b}}$}}
\newcommand{\tauav}{\mbox{$\rm\tau_{av}$}}
\newcommand{\qt}{\mbox{$Q_{\rm T}$}}
\newcommand{\qm}{\mbox{$Q_{\rm M}$}}
\newcommand{\qe}{\mbox{$Q_2$}}
\newcommand{\qtone}{\mbox{$Q_{\rm T1}$}}
\newcommand{\qttwo}{\mbox{$Q_{\rm T2}$}}
\newcommand{\qmtwo}{\mbox{$Q_{\rm M2}$}}
\newcommand{\xitt}{\mbox{$\xi_{\rm TT}$}}
\newcommand{\xitm}{\mbox{$\xi_{\rm TM}$}}
\newcommand{\qjet}{\mbox{$Q_{\rm jet}$}}
\newcommand{\qvtx}{\mbox{$Q_{\rm vtx}$}}
\newcommand{\sqvtx}{\mbox{$\sigma_{Q_{\rm vtx}}$}}
\newcommand{\vqvtx}{\mbox{$\sigma^2_{Q_{\rm vtx}}$}}
\newcommand{\ql}{\mbox{$Q_\ell$}}
\newcommand{\lex}{\mbox{$L_{\rm ex}$}}
\newcommand{\epsb}{\mbox{$\epsilon_{\rm B}$}}
\newcommand{\repsb}{\mbox{$\rm Re(\epsilon_B)$}}
\newcommand{\deltab}{\mbox{$(\Delta\tau/\tau)_{\rm b}$}}
\newcommand{\deltabbg}{\left(\frac{\Delta\tau}{\tau}\right)_{\rm b}}
\newcommand{\acp}{\mbox{$a_{\rm cp}$}}
\newcommand{\ccp}{\mbox{$c_{\rm cp}$}}
\newcommand{\dmd}{\mbox{$\Delta m_{\rm d}$}}
\newcommand{\dms}{\mbox{$\Delta m_{\rm s}$}}
\newcommand{\aiobs}{\mbox{$A^{\rm obs}_i$}}
\newcommand{\fbzero}{\mbox{$f_{\rm B^0}$}}
\newcommand{\fdel}{\mbox{$f_\delta$}}
\newcommand{\nhad}{\mbox{$N_{\rm had}$}}

\newcommand{\meanxe}{\mbox{$\langle x_E\rangle$}}
\newcommand{\dedx}{{\rm d}E/{\rm d}x}
\newcommand{\ltime}{{\cal L}^{\rm time}}
\newcommand{\ltag}{{\cal L}^{\rm tag}}
\newcommand{\pst}{P^T_s}
\newcommand{\psq}{P^Q_s}
\newcommand{\fsj}{f^j_s}
\newcommand{\fbg}{\mbox{$f_{\rm bg}$}}
\newcommand{\fzbs}{f^0_{\rm B_s}}
\newcommand{\fzbbary}{f^0_{\Lambda_{\rm b}}}
\newcommand{\PLB}[3] {Phys.~Lett.\ {B#1} (#2) #3}
\newcommand{\PRL}[3] {Phys.~Rev.\ {Lett.~#1} (#2) #3}
\newcommand{\PRD}[3] {Phys.~Rev.\ {D#1} (#2) #3}
\newcommand{\NIM}[3] {Nucl.~Instrum.\ {Methods~#1} (#2) #3}
\newcommand{\NPB}[3] {Nucl.~Phys.\ {B#1} (#2) #3}
\newcommand{\CPC}[3] {Comp.~Phys.\ {Comm.~#1} (#2) #3}
\newcommand{\ZPC}[3] {Z.~Phys.\ {C#1} (#2) #3}

\newcommand{\EPJ}[3] {Eur.~Phys.\ J.\ {C#1} (#2) #3}
\newcommand{\etal} {et~al.}
\input epsf
\newcommand{\epostfig}[3]{
\begin{figure}[tbp]
\setlength{\epsfxsize}{1.1\hsize}
\hspace*{-0.05\hsize} \epsfbox{#1}
\caption{\label{#2}#3}
\end{figure}
}
\newcommand{\nhadthree}{2\,390\,221}
\newcommand{\nhadtwo}{754\,372}
\newcommand{\ntmlife}{10\,532}
\newcommand{\ntmthree}{293\,416}
\newcommand{\ntmtwo}{100\,703}
\newcommand{\ntmtot}{394\,119}
\newcommand{\tpval}{1.643}
\newcommand{\tzval}{1.523}
\newcommand{\trval}{1.079}
\newcommand{\tpstat}{0.037}
\newcommand{\tzstat}{0.057}
\newcommand{\trstat}{0.064}
\newcommand{\tpsyst}{0.025}
\newcommand{\tzsyst}{0.053}
\newcommand{\trsyst}{0.041}
\newcommand{\acpval}{0.005}
\newcommand{\acpstat}{0.055}
\newcommand{\acpsyst}{0.013}
\newcommand{\acptval}{0.002}
\newcommand{\acptstat}{0.055}
\newcommand{\ccpval}{0.026}
\newcommand{\ccpstat}{0.027}
\newcommand{\ccpsyst}{0.015}
\newcommand{\epsbval}{0.001}
\newcommand{\epsbstat}{0.014}
\newcommand{\epsbsyst}{0.003}
\newcommand{\delbval}{-0.001}
\newcommand{\delbstat}{0.012}
\newcommand{\delbsyst}{0.008}
\newcommand{\tauavval}{1.500}
\newcommand{\tauavstat}{0.003}
\begin{document}
\begin{titlepage}
{\center\Large

EUROPEAN LABORATORY FOR PARTICLE PHYSICS \\

}
\bigskip

{\flushright
CERN-EP/98-195 \\
December 10, 1998 \\
}
\begin{center}
    \LARGE\bf\boldmath
    Measurement of the \bplus\ and \bzero\ Lifetimes \\
    and Search for CP(T) Violation \\
    using Reconstructed Secondary Vertices
\end{center}
\vspace{5mm}
\bigskip

\begin{center}
\Large The OPAL Collaboration \\
\bigskip
\large
%
%
\end{center}
\vspace{5mm}

\bigskip
\begin{abstract}
The lifetimes of the \bplus\ and \bzero\ mesons, and their ratio, 
have been measured in the OPAL experiment using 2.4 million
hadronic \zb\ decays recorded at LEP.
$\zb\rightarrow\bbbar$ decays were tagged
using displaced secondary vertices and high momentum electrons and muons.
The lifetimes were then measured using 
well-reconstructed charged and neutral secondary vertices selected
in this tagged data sample. The results are
\begin{eqnarray*}
\taubp & = & \tpval \pm \tpstat \pm \tpsyst\rm\,ps \\
\taubz & = & \tzval \pm \tzstat \pm \tzsyst\rm\,ps \\
\taubr & = & \trval \pm \trstat \pm \trsyst \,,
\end{eqnarray*}
where in each case the first error is statistical and the second systematic.

A larger data sample of 3.1 million hadronic \zb\ decays 
has been used to search for CP and CPT
violating effects by comparison of inclusive b and \bqbar\ hadron
decays. No evidence for such
effects is seen. The CP violation parameter \repsb\ is measured to
be
\[
\repsb = \epsbval \pm \epsbstat \pm \epsbsyst
\]
and the fractional 
difference between b and \bqbar\ hadron lifetimes is measured to be
\[
\deltabbg \equiv \frac{\rm\tau (b\ hadron) - \tau (\bqbar\ hadron)}
{\rm\tau (average)} = \delbval \pm \delbstat \pm \delbsyst \,.
\]
\end{abstract}

\vspace{5mm}     

\begin{center}
\large
Submitted to Eur.\ Phys.\ J.\ C.
\vspace{9mm}


\end{center}

\end{titlepage}

\begin{center}{\Large        The OPAL Collaboration
}\end{center}\bigskip
\begin{center}{
G.\thinspace Abbiendi$^{  2}$,
K.\thinspace Ackerstaff$^{  8}$,
G.\thinspace Alexander$^{ 23}$,
J.\thinspace Allison$^{ 16}$,
N.\thinspace Altekamp$^{  5}$,
K.J.\thinspace Anderson$^{  9}$,
S.\thinspace Anderson$^{ 12}$,
S.\thinspace Arcelli$^{ 17}$,
S.\thinspace Asai$^{ 24}$,
S.F.\thinspace Ashby$^{  1}$
% ...
}\end{center}\bigskip
\bigskip
$^{  1}$School of Physics and Astronomy, University of Birmingham,
Birmingham B15 2TT, UK
\newline
$^{  2}$Dipartimento di Fisica dell' Universit\`a di Bologna and INFN,
I-40126 Bologna, Italy
\newline
$^{  3}$Physikalisches Institut, Universit\"at Bonn,
D-53115 Bonn, Germany
\newline
$^{  4}$Department of Physics, University of California,
Riverside CA 92521, USA
\newline
$^{  5}$Cavendish Laboratory, Cambridge CB3 0HE, UK
\newline
$^{  6}$Ottawa-Carleton Institute for Physics,
Department of Physics, Carleton University,
Ottawa, Ontario K1S 5B6, Canada
\newline
$^{  7}$Centre for Research in Particle Physics,
Carleton University, Ottawa, Ontario K1S 5B6, Canada
\newline
$^{  8}$CERN, European Organisation for Particle Physics,
CH-1211 Geneva 23, Switzerland
\newline
$^{  9}$Enrico Fermi Institute and Department of Physics,
University of Chicago, Chicago IL 60637, USA
\newline
$^{ 10}$Fakult\"at f\"ur Physik, Albert Ludwigs Universit\"at,
D-79104 Freiburg, Germany
\newline
$^{ 11}$Physikalisches Institut, Universit\"at
Heidelberg, D-69120 Heidelberg, Germany
\newline
$^{ 12}$Indiana University, Department of Physics,
Swain Hall West 117, Bloomington IN 47405, USA
\newline
$^{ 13}$Queen Mary and Westfield College, University of London,
London E1 4NS, UK
\newline
$^{ 14}$Technische Hochschule Aachen, III Physikalisches Institut,
Sommerfeldstrasse 26-28, D-52056 Aachen, Germany
\newline
$^{ 15}$University College London, London WC1E 6BT, UK
\newline
$^{ 16}$Department of Physics, Schuster Laboratory, The University,
Manchester M13 9PL, UK
\newline
$^{ 17}$Department of Physics, University of Maryland,
College Park, MD 20742, USA
\newline
$^{ 18}$Laboratoire de Physique Nucl\'eaire, Universit\'e de Montr\'eal,
Montr\'eal, Quebec H3C 3J7, Canada
\newline
$^{ 19}$University of Oregon, Department of Physics, Eugene
OR 97403, USA
\newline
$^{ 20}$CLRC Rutherford Appleton Laboratory, Chilton,
Didcot, Oxfordshire OX11 0QX, UK
\newline
$^{ 22}$Department of Physics, Technion-Israel Institute of
Technology, Haifa 32000, Israel
\newline
$^{ 23}$Department of Physics and Astronomy, Tel Aviv University,
Tel Aviv 69978, Israel
\newline
$^{ 24}$International Centre for Elementary Particle Physics and
Department of Physics, University of Tokyo, Tokyo 113-0033, and
Kobe University, Kobe 657-8501, Japan
\newline
$^{ 25}$Institute of Physical and Environmental Sciences,
Brunel University, Uxbridge, Middlesex UB8 3PH, UK
\newline
$^{ 26}$Particle Physics Department, Weizmann Institute of Science,
Rehovot 76100, Israel
\newline
$^{ 27}$Universit\"at Hamburg/DESY, II Institut f\"ur Experimental
Physik, Notkestrasse 85, D-22607 Hamburg, Germany
\newline
$^{ 28}$University of Victoria, Department of Physics, P O Box 3055,
Victoria BC V8W 3P6, Canada
\newline
$^{ 29}$University of British Columbia, Department of Physics,
Vancouver BC V6T 1Z1, Canada
\newline
$^{ 30}$University of Alberta,  Department of Physics,
Edmonton AB T6G 2J1, Canada
\newline
$^{ 31}$Duke University, Dept of Physics,
Durham, NC 27708-0305, USA
\newline
$^{ 32}$Research Institute for Particle and Nuclear Physics,
H-1525 Budapest, P O  Box 49, Hungary
\newline
$^{ 33}$Institute of Nuclear Research,
H-4001 Debrecen, P O  Box 51, Hungary
\newline
$^{ 34}$Ludwigs-Maximilians-Universit\"at M\"unchen,
Sektion Physik, Am Coulombwall 1, D-85748 Garching, Germany
\newline
\bigskip\newline
$^{  a}$ and at TRIUMF, Vancouver, Canada V6T 2A3
\newline
$^{  b}$ and Royal Society University Research Fellow
\newline
$^{  c}$ and Institute of Nuclear Research, Debrecen, Hungary
\newline
$^{  d}$ and Department of Experimental Physics, Lajos Kossuth
University, Debrecen, Hungary
\newline
$^{  e}$ on leave of absence from the University of Freiburg
\newpage
\section{Introduction}

The lifetimes of b hadrons depend both on the strength of the b quark 
coupling to the lighter c and u quarks, and on the dynamics of b hadron
decay. In the spectator model of heavy hadron decay, the decay of the
heavy quark is unaffected by the presence of the other light 
quarks in the hadron, so the lifetimes of all hadrons containing
the same heavy quark are predicted to be
equal. This model fails badly for the charm hadrons, where the
lifetime of the $\rm D^+$ is more than 2.5 times the lifetime of the
$\rm D^0$ \cite{pdg98}, but is a better approximation
for the b hadrons, due to the larger mass of the b quark \cite{spec}.
The difference between the \bplus\ and \bzero\ 
lifetimes\footnote{Charge conjugate states are implied when 
discussing the lifetimes of individual b hadron species.} is expected
to be at most $10\,\%$, and depends on the details of the various
non-spectator processes contributing to their decay. Measurements at
the level of a few percent or better are therefore needed to test this
prediction and probe the dynamics of b hadron decays.

Experimentally, most measurements of the \bplus\ and \bzero\ lifetimes
have been performed using semileptonic decays, fully or partially
reconstructing the decay products to distinguish \bplus\ from
\bzero\ \cite{opalsemil,bsemil}. These measurements are limited 
due to the small
branching ratios and limited reconstruction efficiencies for the
selected final states. A more inclusive approach is to reconstruct
resolvable secondary vertices from b hadron decays, since their long
lifetimes lead to significant decay lengths \cite{delphitop,bvtx}. 
In these analyses, the
\bplus\ and \bzero\ decays are distinguished by reconstructing the
charge of the secondary vertex. This technique results in much larger
data samples, and offers the best hope of improving the
precision. A measurement of this type is presented in the first part
of this paper.

Inclusive samples of \bzero\ decays can also be used to search
for CP and CPT violating effects. Although CP violation has so far
been seen only in the neutral kaon system,  possibly large effects are
expected also in the B-meson system, so it is worthwhile to
search for them even with the relatively small data samples collected at LEP.
In the \bzero-\bzerobar\ system the weak eigenstates $|\bzero\rangle$
and $|\bzerobar\rangle$
differ from the mass eigenstates $|\bx\rangle$ and $|\by\rangle$:
\begin{eqnarray}\label{e:eigen}
|\bx\rangle & = & \frac{(1+\epsb+\delta_B)|\bzero\rangle
+(1-\epsb-\delta_B)|\bzerobar\rangle}
{\sqrt{2 (1+|\epsb+\delta_B|^2)}}  \\
|\by\rangle & = & \frac{(1+\epsb-\delta_B)|\bzero\rangle
-(1-\epsb+\delta_B)|\bzerobar\rangle}
{\sqrt{2 (1+|\epsb-\delta_B|^2)}} \nonumber
\end{eqnarray}
where the parameters \epsb\ and $\delta_B$ parameterise indirect
CP violation and CPT violation respectively \cite{rebimb}. 
These parameters can be studied using semi-leptonic b hadron decays,
and limits of a few
$10^{-2}$ have been set on both quantities \cite{rvk,oldeb,opaldms,rjh}. 
A non-zero value of \epsb\ is also expected to produce
time dependent asymmetries in inclusive \bzero\ {\em vs.\/} inclusive
\bzerobar\ decays \cite{cpinc}. This provides a second method to look
for CP violating effects in b decays, and such an asymmetry is
searched for in the second part of this paper. The same data sample is also
used to test a basic prediction of CPT invariance, that the lifetimes
of b and \bqbar\ hadrons are equal.

A brief overview of the analysis
strategies is presented in Section~\ref{s:over}, followed by a review of
the data and Monte Carlo samples in Section~\ref{s:datamc}. The parts
common to both \bplus\ and \bzero\ lifetime and CP(T) violation
analyses, namely
the \bbbar\ event tagging and b hadron production flavour tagging, are
discussed in Sections~\ref{s:btag} and~\ref{s:bprod}. The lifetime
analysis is described in detail in Section~\ref{s:blife} and the 
CP and CPT violation analyses in Section~\ref{s:cpt}. Finally, all the
results and conclusions are summarised in Section~\ref{s:conc}.

\section{Analysis Overview}\label{s:over}

The analyses exploit the characteristic topology of the 
$\zb\rightarrow\bbbar$ decay: two back-to-back jets aligned along the 
direction of the thrust axis.
The event is divided into two hemispheres by the plane perpendicular 
to the thrust axis and containing the $\rm e^+e^-$ interaction point.
One hemisphere (the `tag hemisphere' T) 
is tagged as containing a b decay using either a 
displaced secondary vertex or a high momentum lepton. The production
flavour of the b hadron in the tag hemisphere (whether it originated
from a b quark or $\rm\overline{b}$ antiquark) is also determined,
using jet, vertex and lepton charge information.
The unbiased b decay in the other hemisphere (the `measurement hemisphere'
M) is used to perform the measurement of b hadron decay time.

The decay time of the b hadron in the measurement hemisphere is determined
by reconstructing its decay vertex and energy. In the \bplus\ and \bzero\ 
lifetime analysis, this decay vertex is required to be significantly
separated from the event primary vertex, and very
strict requirements are placed on the confidence with which tracks are 
assigned to either this secondary vertex or the primary vertex. These
requirements lead to a final data sample of only about 10\,000 reconstructed
vertices with well determined charge.
The reconstructed decay times of the b hadrons giving rise to these
charged and neutral vertices are then used
to determine the lifetimes of the \bplus\ and \bzero\ mesons, employing
the excess decay length technique~\cite{delphitop,na14} to eliminate biases
caused by the separated vertex requirement. The correlation of the
sign of reconstructed charged b hadrons in the measurement hemisphere with
the production flavour of the b hadron in the tag hemisphere is
used to measure the probability of mis-reconstructing the b hadron charge.

In the CP(T) violation analysis, only the decay time of the b hadron
in the measurement hemisphere is reconstructed. The charge is not
determined, so no strict requirements are placed on the quality or separation
of the b hadron vertex, leading to a much larger data sample of 
about 400\,000 events.
The production flavour of this b hadron is inferred from that of the b hadron
in the tag hemisphere, aided by information 
in the measurement hemisphere, and taking into account the effects of
\bzero\ and \bs\ mixing. The decay time distribution of decays 
tagged as b or $\rm\overline{b}$ hadrons is then used to search for
CP and CPT violating effects.

\section{Data sample and event simulation}\label{s:datamc}

The OPAL detector is well described elsewhere 
\cite{opaldet,opalsi2d,opalsi3d}. The
analyses described here rely mainly on charged particle track
reconstruction using the central tracking chambers and the silicon
microvertex detector. The b hadron lifetime analysis requires 
excellent pattern recognition, and uses only data taken between
1993 and 1995 with the upgraded silicon microvertex detector 
measuring tracks in both the $r$-$\phi$ and $r$-$z$ 
planes\,\footnote{A right handed coordinate system is used, with
positive $z$ along the $\mathrm{e}^-$ beam direction and $x$ pointing
towards the centre of the LEP ring. The polar and azimuthal angles are
denoted by $\theta$ and $\phi$, and 
the origin is taken to be the centre of the detector.}
\cite{opalsi3d}. The CP(T) violation analysis also uses
data taken in 1991 and 1992 with the original silicon microvertex 
detector measuring only in the $r$-$\phi$ plane \cite{opalsi2d}.

Hadronic \zb\ decays were selected using standard criteria 
\cite{evtsel}, additionally requiring at least 7~charged tracks in the
event. The thrust axis polar angle 
$\cos\theta_T$ was calculated using charged tracks and
electromagnetic clusters not associated to any track. To ensure the
event was well contained within the acceptance of the detector, the
thrust axis direction was required to satisfy $|\cos\theta_T|<0.9$.
The complete event selection has an efficiency of about 88\,\%, and
selected a total of \nhadthree\ events in the 1993--95 data and 
\nhadtwo\ events in the 1991-92 data.

Charged tracks and electromagnetic calorimeter clusters with no
associated track were combined into jets using a cone
algorithm~\cite{jetcone}, with a cone half angle of 0.65\,rad and
a minimum jet energy of 5\,GeV. The transverse momentum $p_t$ of each
track was defined relative to the axis of the jet containing it, where
the jet axis was calculated including the momentum of the track.

Monte Carlo simulated events were generated using JETSET 7.4
\cite{jetset} with parameters tuned by OPAL \cite{jetsetopt}. The
fragmentation functions of Peterson et al.~\cite{fpeter} were used to
describe the fragmentation of b and c quarks. The generated events
were passed through a program that simulated the response of the OPAL
detector \cite{gopal} and through the same reconstruction algorithms 
as the data.

\section{Tagging $\rm\bf b\overline{b}$ events}\label{s:btag}

Two methods were used to tag \bbbar\ events, based on
displaced secondary vertices
and high momentum leptons. The first method exploits the long
lifetime, hard fragmentation, 
high decay multiplicity and high mass of b hadrons. An attempt was made to 
reconstruct a significantly separated secondary vertex in each jet of the 
event. If a secondary vertex was found, an  artificial neural network 
was used to further separate 
b decays from charm and light quark background. The neural
network has five inputs, derived from decay length, vertex multiplicity and
invariant mass information. The algorithm is
described fully in \cite{opalrb}. For this analysis, a hemisphere was
defined to be tagged if any jet in the hemisphere had a secondary vertex
with tag variable $B>1.2$ \cite{opalrb}, 
giving a hemisphere tagging efficiency 
of 32\,\% in \bbbar\ events, and a non-b impurity of 12\,\%.

Electrons and muons with momentum $p>2\rm\,GeV$ and transverse momentum 
$p_t>1\rm\,GeV$ were also used to tag \bbbar\ events\,\footnote{The notation
$c=1$ is employed in this paper.}.
Electrons were
identified in the polar angle region $|\cos\theta|<0.9$ using a
re-optimised version of the neural network algorithm described in
\cite{elecid}. The identification relies on ionisation energy loss
($\dedx$) measured in the tracking chamber, 
spatial and energy-momentum ($E/p$) matching between tracking and
calorimetry, and the multiplicity measured in the presampler
detectors. Photon conversions were rejected using another neural
network algorithm \cite{elecid}. Muons were identified in the polar
angle region $|\cos\theta|<0.9$ by requiring a spatial match between
a track reconstructed in the tracking detectors and a track segment
reconstructed in the external muon chambers, as in \cite{muonid}.

The tagged lepton hemispheres were further enhanced in semileptonic b
decays by using information from the lepton $p$ and $p_t$, and its degree
of isolation from the rest of the jet, in a neural network
algorithm \cite{opaldil}. This suppresses contributions from
cascade ($\rm b\rightarrow c\rightarrow \ell$) decays (which have the
wrong correlation of lepton sign with b hadron production flavour),
charm and fake lepton background. 
The output $S$ of the network was required to be greater than 0.7,
giving a lepton sample about 75\,\% pure in $\rm b\rightarrow\ell$ decays.
The lepton tags contribute an additional 5\,\% to the hemisphere b-tagging
efficiency, bringing the total
to about 37\,\% with an impurity of 13\,\% in Monte Carlo.

The events used for the final analysis are those tagged by either of
the b tagging methods described above (referred to as the T-tag)
in one hemisphere, and passing the measurement selection requirements
in the other hemisphere. These latter requirements are described below
in Section~\ref{ss:lfvtx} for the b hadron lifetime analysis and 
Section~\ref{ss:cpvtx} for the CP(T) analysis, and are referred to as
the M-tag. Both hemispheres are used as measurement hemispheres in 
events tagged by both tags in both hemispheres.

The \bbbar\ purity of the combined tag T-M samples (tagged
by the T-tag in one hemisphere and the M-tag in the other) were
determined by applying an extension of the double tagging technique
used for measuring \rb\ \cite{opalrb}. The number of hemispheres $N_i$
tagged by tag $i$ ($i$=T or M), and the number of events $N_{ij}$
tagged by tag $i$ in one hemisphere and tag $j$ in the other
hemisphere, are related to the total number \nhad\ of events passing
the event selection by:
\begin{eqnarray}
N_i & = & 2 \nhad \{ \epsilon_i^{\rm b}\ \rb\ +\epsilon_i^{\rm c}\
\rc\ +\epsilon_i^{\rm uds}\ (1-\rb - \rc ) \} ,
\label{e:ntmult} \\
N_{ij} & = & (2-\delta_{ij}) \nhad \{ 
C_{ij}^{\rm b}\ \epsilon_i^{\rm b} \epsilon_j^{\rm b}\ \rb +
C_{ij}^{\rm c}\ \epsilon_i^{\rm c} \epsilon_j^{\rm c}\ \rc +
C_{ij}^{\rm uds}\ \epsilon_i^{\rm uds} \epsilon_j^{\rm uds}\
(1-\rb-\rc )\} . \nonumber
\end{eqnarray}
Here $\epsilon_i^{\rm q}$ gives the efficiency of tag $i$ to tag
hemispheres of flavour q (q=uds, c or b). The correlations
$C_{ij}^{\rm q}$ are defined by 
$C_{ij}^{\rm q}=\epsilon_{ij}^{\rm q}/(\epsilon_i^{\rm q}\epsilon_j^{\rm
q})$ where $\epsilon_{ij}^{\rm q}$ is the efficiency to tag a \qqbar\
event simultaneously with tag $i$ in one hemisphere and tag $j$ in the
other. Deviations of $C_{ij}^{\rm q}$ from unity account for the fact
that the tagging in the two hemispheres is not completely independent,
there being small efficiency correlations for both physical and
instrumental reasons.
The quantities \rb\ and \rc\ are the fractions of hadronic \zb\ decays
to \bbbar\ and \ccbar, and were taken to be $\rb=0.2170\pm 0.0009$ 
and $\rc=0.173\pm 0.005$ \cite{pdg98}.

The values of \ect\ and \eudst\ (which are known to be well 
modelled in Monte Carlo \cite{opalrb}) together with all the correlation
terms $C_{ij}^{\rm q}$, were determined from the Monte Carlo, and the values of
\nhad, $N_i$ and $N_{ij}$ measured from the data. The five 
equations~\ref{e:ntmult} were then solved using a $\chi^2$ minimisation
procedure to give the b-tagging efficiency \ebt\ of the T-tag, and all
the tagging efficiencies \ebm, \ecm\ and \eudsm\ of the M-tag. The
b-purity \tmpure\ of the combined tag T-M sample was then calculated as:
\begin{equation}
\tmpure=\frac{C_{\rm TM}^{\rm b}\ \ebt \ebm\ \rb}
{C_{\rm TM}^{\rm b}\ \ebt \ebm\ \rb +
C_{\rm TM}^{\rm c}\ \ect \ecm\ \rc +
C_{\rm TM}^{\rm uds}\ \eudst \eudsm\
(1-\rb-\rc )} .
\end{equation}
The results of this procedure are given in Table~\ref{t:purity} for
the three data samples: 1993--95 data with the M-tag used to measure the 
b hadron lifetime, and 1993--95 and 1991-92 data with the
M-tag used in the CP(T) analysis. 
In the last case, the T-tag used vertexing in the
$r$-$\phi$ plane only \cite{opalrb}, and the purity is correspondingly
lower. The purity is
somewhat higher for the b lifetime analysis, as the requirements on
the M-tag secondary vertex also provide some rejection of non-b
background. In contrast, the CP(T) M-tag has almost equal efficiency for
all flavours.

\begin{table}
\centering

\begin{tabular}{l|rrr}\hline\hline
Data Sample & 1993--95 & 1993--95 & 1991--92 \\
&  b lifetime & CP(T) & CP(T)  \\
\hline
Number of events & \nhadthree & \nhadthree & \nhadtwo  \\
Number of T-M events & \ntmlife & \ntmthree & \ntmtwo \\
\hline
Combined tag purity (\%) & $94.8\pm 1.5$ & $87.9\pm 2.6$ & $81.8\pm 3.6$ \\
Statistical error & 0.7 & 0.4 & 0.7 \\
Systematic errors: & & & \\
\ \ T-tag uds efficiency & 0.1 & 1.3 & 1.7 \\
\ \ T-tag charm efficiency & 0.3 & 0.4 & 0.7 \\
\ \ Correlations (udsc events) & 1.2 & 2.2 & 3.0 \\
\ \ Correlations (b events) & 0.5 & 0.4 & 0.4 \\
\ \ \rc\ value              & 0.1 & 0.2 & 0.3 \\
Total systematic error & 1.3 & 2.6 & 3.5 \\
\hline
Total error (\%) & 1.5 & 2.6 & 3.6 \\
\hline
\end{tabular}
\caption{\label{t:purity} The numbers of hadronic events, selected 
combined tag T-M events and tag purities for each of the data samples.
The breakdown of statistical and systematic errors for each of the
purity values is also given.}
\end{table}

The systematic errors resulting from each of the inputs used in the fits for
the b purity are also given in Table~\ref{t:purity}. They were
evaluated using the methods and parameter ranges discussed in
\cite{opalrb}. The efficiency errors include the effects of charm and
light quark physics uncertainties, tracking resolution and lepton 
identification. The hemisphere tagging correlations for \bbbar\ events
are slightly larger than those in \cite{opalrb}, and a 
systematic uncertainty of $\pm 0.02$ on the $C^{\rm b}_{ij}$ values 
is estimated, in addition to the Monte Carlo
statistical errors. The correlations are larger because of stronger
geometrical effects at large values of $|\cos\theta_T|$. The 
uncertainties on the uds and \ccbar\ correlations 
$C^{\rm uds}_{ij}$ and $C^{\rm c}_{ij}$ include
systematic errors of $\pm 0.2$ 
but are still dominated by Monte Carlo  statistics.
The error from uncertainty in the value of \rb\ is negligible.

\section{Tagging the b production flavour}\label{s:bprod}

The production flavour (b or \bqbar) of the b hadron in the T-tagged
hemisphere was determined using up to three pieces of information in
each event: the momentum-weighted average track charge, or `jet charge';
the charge of a  secondary vertex reconstructed in the hemisphere; and
the charge of a high momentum lepton in the hemisphere. The jet charge
can be calculated for every tagged hemisphere, and either or both of  
the vertex or lepton charges is always available as a consequence of 
the b-tagging techniques used. The three charges were combined using a 
neural network algorithm to produce a single production flavour tag
 variable \qt\ for each T-tagged hemisphere.

The jet charge \qjet\ was calculated for the highest energy jet in the T-tag 
hemisphere as:
\begin{equation}
\qjet=\frac{\sum_i (p_i^l)^\kappa q_i}{\sum_i (p_i^l)^\kappa}\label{e:qjet}
\end{equation}
where $p_i^l$ is the longitudinal momentum component with respect
to the jet axis and $q_i$ the charge ($\pm 1$) of track
$i$, and the sum was taken over all the tracks in the jet. The
parameter $\kappa$ was set to 0.5 to optimise the separation between 
hemispheres containing b and \bqbar\ hadrons, including the effects of
\bzero\ and \bs\ mixing \cite{opaljpks}.

For hemispheres tagged by a secondary vertex, the charge of this
vertex \qvtx\ was calculated as:
\begin{equation}
\qvtx=\sum_i w_i q_i\label{e:qvtx}
\end{equation}
and the uncertainty \sqvtx\ as:
\begin{equation}
\vqvtx=\sum_i w_i (1-w_i) q^2_i\label{e:eqvtx}
\end{equation}
where $w_i$ is the weight for track $i$ to have come from the
secondary, rather than the primary, vertex \cite{opalbstar}.
The weights $w_i$ were
obtained from a neural network algorithm using as input the track momentum,
transverse momentum with respect to the jet axis, and impact
parameters with respect to the reconstructed primary and secondary
vertices, as in \cite{opaljpks}. A well reconstructed vertex charge 
(with small \sqvtx) close
to $+1$ ($-1$) indicates a \bplus\ (\bminus), tagging the hemisphere as
containing a \bqbar\ (b) quark, whilst a vertex charge close to zero
indicates a neutral b hadron ({\em e.g.\/} \bzero\ or \bzerobar), giving no
information on the production flavour. A vertex charge with large
\sqvtx\ cannot distinguish between $\qvtx=0$ or $\qvtx=\pm 1$, so
again provides no information on the production flavour.

For hemispheres tagged by a lepton, the lepton charge gives an
estimate of the b production flavour, diluted by \bzero\ and \bs\
mixing, cascade b decays,
charm decays and fake leptons. The product $\ql S$ of the lepton charge
\ql\ and the output $S$ of the neural network used to select 
$b\rightarrow\ell$ decays (see Section~\ref{s:btag})
was used as a tagging variable
analogous to \qjet\ and \qvtx. Higher values of $S$ indicate greater
probability of the lepton originating from a semileptonic b decay.

The available production flavour estimates for each T-tagged
hemisphere were combined into a single estimate \qt\ using a neural
network algorithm, as in \cite{opaljpks}. The neural network has four
inputs: the jet charge \qjet, the vertex charge \qvtx\ and its error 
\sqvtx, and the lepton tag $\ql S$. Separate neural networks with
fewer inputs were trained for use in hemispheres with only a
vertex or a lepton. The variable \qt\ is defined such that:
\[
\qt=\frac{N_{\rm b}(x)-N_{\rm\bar{b}}(x)}{N_{\rm b}(x)+N_{\rm\bar{b}}(x)}
\]
where $N_{\rm b}(x)$ and $N_{\rm\bar{b}}(x)$ are the numbers of Monte 
Carlo b hadron and \bqbar\ hadron hemispheres
with a particular value of the neural network output $x$.
Thus hemispheres with $\qt=+1$ are tagged with
complete confidence as b hadrons, hemispheres with $\qt=-1$ with 
complete confidence as \bqbar\ hadrons, and hemispheres with $\qt=0$
are equally likely to be b or \bqbar\ hadrons. The modulus $|\qt|$ 
satisfies
$|\qt|=1-2\eta$, where $\eta$ is the `mis-tag probability', {\em i.e.\/}
the probability to tag the production flavour incorrectly.

For the CP(T) violation analysis, both the production flavour estimate
from the T-tag hemisphere and the jet charge \qjet\ in the M-tag
hemisphere are used to infer the production flavour of the b hadron in the
M-tag hemisphere.
In this hemisphere, the parameter $\kappa$ in
equation~\ref{e:qjet} is set to zero, so the jet charge becomes simply
the average of the charges of the tracks in the jet. This avoids being
sensitive to the decay flavour of mixed or unmixed \bzero\ and \bs\
mesons, but is still sensitive to the production flavour of the b
hadron via the information carried by the fragmentation tracks in the
jet \cite{opaldstar}. This jet charge is used to produce a tagging
variable \qm\ defined similarly to \qt\ for the T-tag hemisphere. The
information from both hemispheres is combined into a single variable
\qe, defined as:
\[
\qe=2\left\{\frac{(1-\qt)(1+\qm)}{(1-\qt)(1+\qm)+(1+\qt)(1-\qm)}\right\}-1
\]
The variable \qe\ is positive (negative) if the event is tagged as 
containing a b (\bqbar) hadron in the M-tag hemisphere.

The \qjet\ and \qvtx\ distributions are not charge symmetric because
of detector effects causing a difference in the rate and
reconstruction of positive and negative tracks. These effects are
caused by the detector material and the Lorentz angle in the tracking
chambers \cite{opaldms}.
They were removed by subtracting offsets from the \qjet\ and
\qvtx\ values before the calculation of \qt\ and \qm. 
The offsets were determined
from data using the inclusive T-tagged samples, and were found to be
fractions $0.018\pm 0.002$, $0.034\pm 0.002$ and $0.028\pm 0.002$ 
of the RMS width of the $\qjet (\kappa=0.5)$,
$\qjet (\kappa=0)$ and \qvtx\ distributions, respectively.

The distributions of \qt\ and \qm\ in T-tagged data and Monte Carlo events 
are shown in Figures~\ref{f:ostag}(a) and~\ref{f:ostag}(b). 
The sharp drop at about $|\qt|=0.8$ in Figure~\ref{f:ostag}(a) is due to
the irreducible fraction of lepton tagged events that are tagged incorrectly
due to \bzero\ and \bs\ mixing. Around
30\,\% of events have equal numbers of positive and negative tracks in
the jet in the M-tag hemisphere, giving zero jet charge and
$\qm=0$. These events are not shown in Figure~\ref{f:ostag}(b).
Some discrepancies between data and Monte Carlo distributions are
visible. These discrepancies are not important, provided that the
\qt\ values in data correctly describe the data mistag probabilities.
To check that this is the case, events where both hemispheres are 
T-tagged, yielding b production flavour estimates \qtone\ and \qttwo, were
used. Since the two hemispheres must contain b hadrons produced
with opposite flavour, 
the product $\xitt=\qtone\qttwo$ is negative if both (or neither)
hemispheres are tagged correctly, and positive if only one is 
tagged incorrectly.
The distribution $f(\xitt)$ of \xitt\ in data thus allows the 
production flavour estimate \qt\ to be checked. 
The function:
\begin{equation}
g(\xitt)\equiv \frac{f(\xitt)}{f(\xitt)+f(-\xitt)}
\ \ \ \{ \mbox{for $\xitt > 0$} \} \label{e:xitt}
\end{equation}
represents the `wrong sign fraction' at a particular value of $|\xitt|$, and 
should satisfy $g(\xitt)=(1-\xitt)/2$ if \qt\  correctly describes
the mis-tag probabilities. The distribution $g(\xitt)$ in data is shown in 
Figure~\ref{f:ostag}(c), together with a linear fit. The distribution has
the expected form, and the fitted gradient is $dg/d\xitt=-0.514\pm 0.012$,
showing that the average magnitude of \qt\ is correct to a relative precision
of 2.6\,\%.

\epostfig{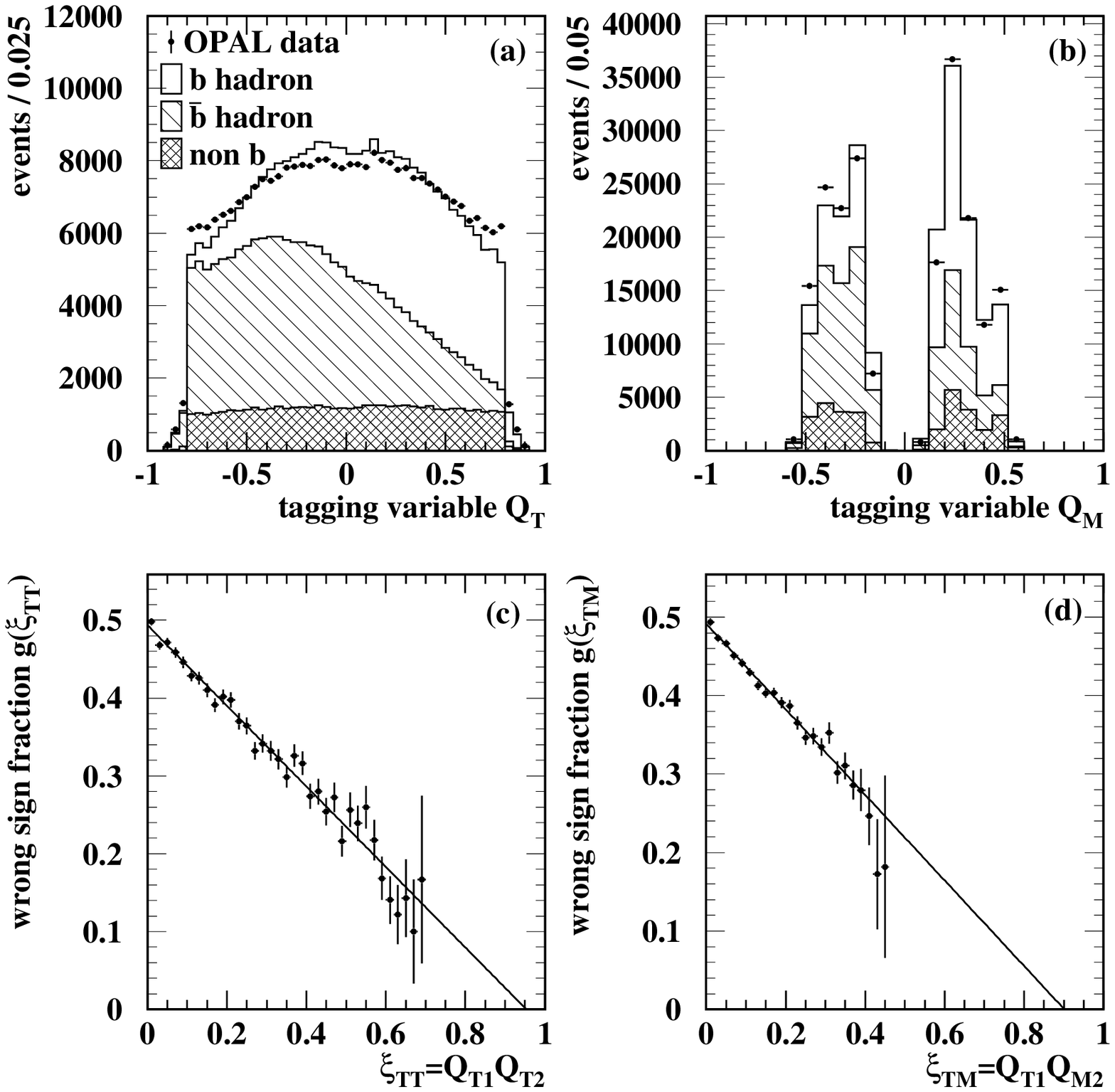}{f:ostag}{Production flavour tagging of \bbbar\
  events: (a) Distribution of the flavour tagging variable \qt\ 
in data (points) and Monte Carlo (histogram) T-tagged hemispheres. The
  contributions from b hadrons, \bqbar\ hadrons and non-b backgrounds
  are shown. (b) The analogous distributions for the M-tag
  hemispheres. The 30\,\% of the sample (90000 events in data)
  with $\qm=0$ are not shown. In both (a) and (b), the error bars on
  the data points are smaller than the symbols.
  (c) Distribution of the wrong sign fraction $g(\xitt)$ in data double
  T-tagged events, with a linear fit superimposed. (d) 
  Distribution of the wrong sign fraction $g(\xitm)$ in data
  double T-tagged events, again with a linear fit superimposed.}

The same technique was used to study \qm, via the `cross-tag' product 
$\xitm=\qtone\qmtwo$, as \qm\ has a smaller tagging power than \qt.
The function $g(\xitm)$ was defined in an analogous
way  to $g(\xitt)$ in  
equation~\ref{e:xitt}. The distribution $g(\xitm)$ in data is shown  
in Figure~\ref{f:ostag}(d), and the corresponding gradient is
$-0.545\pm 0.018$. Since \qt\ has already been shown above to be correct to
2.6\,\%, all of this discrepancy is attributed to \qm, which therefore
has a bias of up to about $10\,\%$. These values are used to
calculate tagging systematic errors.

\section{$\rm\bf B^+$ and $\rm\bf B^0$ lifetime analysis}\label{s:blife}

The \bbbar\ event tagging and production flavour tagging described above
is common to both the b hadron lifetime and CP(T) analyses. 
The remainder of each 
analysis---the M-tag and the fits to the data---are specific to each analysis.
The b hadron lifetime analysis is described in this section, and the 
CP(T) analysis in Section~\ref{s:cpt}.

\subsection{b hadron reconstruction}\label{ss:lfvtx}

The b hadron reconstruction used for the M-tag in this analysis 
is similar to that used in \cite{delphitop}. It aims to
reconstruct a relatively small sample of clear secondary vertices where
each track can be unambiguously associated to either the primary or secondary
vertex. 

The algorithm considers all tracks in a jet which have
momentum $p>0.5$\,GeV, impact parameter (in the $r$-$\phi$ plane) 
$|d_0|<1$\,cm and error on the impact parameter $\sigma_{d_0}<0.1$\,cm.
All possible sets of assignments of these tracks to the primary and
secondary vertices (`arrangements')
were considered, requiring at least two tracks to
be assigned to the secondary vertex, and including the combination where
no tracks at all are assigned to the primary vertex.

The positions of the two vertices in each arrangement were determined
by fitting all the assigned tracks to a common vertex in three dimensions.
For the primary vertex, an additional constraint from the average 
beam spot position was used, determined from a fit to the tracks in 
many consecutive events \cite{beamspot}. The $\chi^2$ of the 
arrangement was calculated as the sum of the $\chi^2$ values for the
primary and secondary vertex fits, and the fit probability of the arrangement
determined from the $\chi^2$  value and the number of degrees of freedom
in the two vertex fits.

For the jet to be accepted as having a clear secondary vertex, the following
conditions had to be satisfied:
\begin{enumerate}
\item One and only one arrangement (the `best arrangement') 
has a fit probability exceeding 1\,\%.
\item All other arrangements have a $\chi^2$ value exceeding that of the
best arrangement by at least 4.
\item The decay length $L$ of the secondary vertex, divided by its
error $\sigma_L$, satisfies $L/\sigma_L>3$. The decay length is calculated
from the distance between the fitted primary and secondary vertices, using
the direction of the jet axis as a constraint \cite{opalrb}. 
\item The decay length must be positive, but less than 3\,cm. The
  decay length is positive if the secondary vertex is displaced from
  the primary vertex in the same direction as the jet momentum vector.
\end{enumerate}
These requirements ensure that the best arrangement has an acceptable 
probability, that no other arrangement is likely to be the correct one,
and that the primary and secondary vertices are well separated.
Jets containing b hadrons
with short decay lengths will tend to fail requirements~2 or~3, whilst
those with a mis-measured track which does not fit with either the  primary
or secondary vertex will tend to fail requirement~1.

The charge $Q$ of the secondary vertex and its error $\sigma_Q$ 
were then calculated, using
equations~\ref{e:qvtx} and~\ref{e:eqvtx}, with track weights $w_i$ 
optimised for this vertex finding algorithm. To ensure the vertex
charge was well reconstructed, the error $\sigma_Q$ was required to be less
than 0.7. All tracks in the jet were used in calculating the vertex
charge, including those which failed the tighter selection used for the
initial vertex finding.

A total of \ntmlife\ combined T-M tagged events were found in the data
with this selection for the M-tag. The distribution of the M-tag 
vertex charge,
together with the Monte Carlo prediction, is shown in Figure~\ref{f:qvtx}.
Clear peaks at $Q=\pm 1$ and 0 are seen, corresponding to high
purities of charged and
neutral b hadrons. In the Monte Carlo, the neutral b hadrons are a mixture of
approximately 63\,\% \bzero, 24\,\% \bs\ and 13\,\% b baryons
(denoted by \bbary). The charged b sample is almost entirely \bplus, with a
very small contribution from charged b baryons ($\Xi_{\rm b}^-$) which is
estimated at about 1\,\% and is neglected.

\epostfig{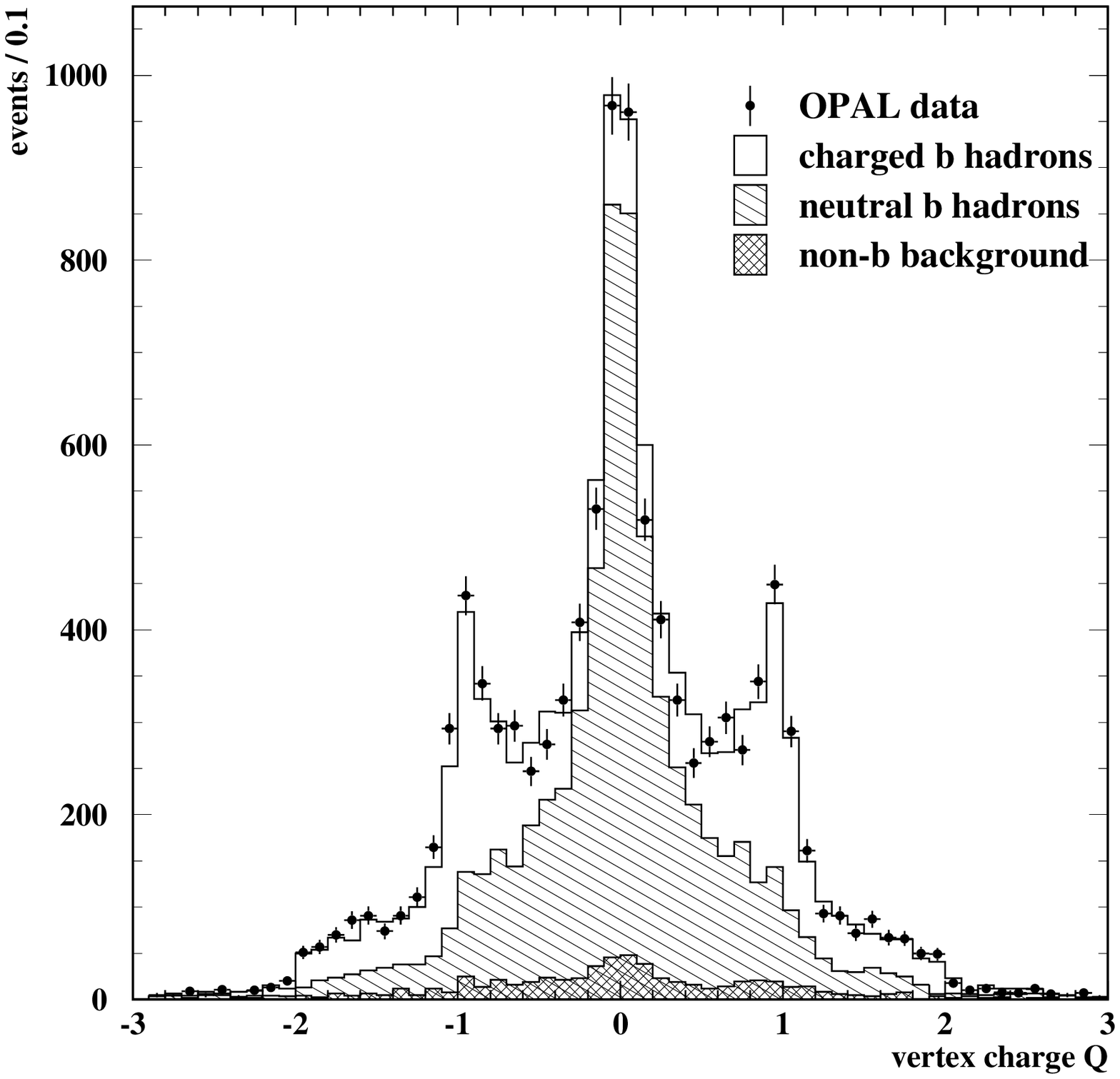}{f:qvtx}{Distributions of the vertex charge $Q$ 
in the data (points) and Monte Carlo (histogram), normalised to the
same number of entries. The expected contributions from charged b
hadrons, neutral b hadrons and non-b background are indicated.}

\subsection{Excess proper time reconstruction}\label{ss:tlex}

The decay length distributions of secondary vertices selected by the
above algorithm in data and Monte Carlo are shown in Figure~\ref{f:decl}(a).
The decay lengths are very biased towards large values, since the
algorithm preferentially selects well separated secondary vertices.
Therefore in order to extract the b hadron lifetimes, the excess decay length
method is used, as in \cite{delphitop,na14}. For each selected event, 
the minimum b hadron decay length that would still result in a
resolvable secondary vertex passing all the requirements described in
Section~\ref{ss:lfvtx} was determined.  To find this minimum decay length,
all the tracks assigned to the
secondary vertex in the best arrangement were assumed to come from the
b hadron. They were then translated along the direction of the jet
axis towards the primary vertex (as if the b hadron had decayed at an
earlier time), and the entire vertex selection repeated (requirements
1--4 of Section~\ref{ss:lfvtx} and $\sigma_Q<0.7$, including
recalculating the $\chi^2$ values for all possible vertex
arrangements). The translation distance at the point where the
modified arrangement just fails one of the requirements defines the excess
decay length \lex. The distributions of \lex\ in data and Monte Carlo 
are shown in Figure~\ref{f:decl}(b).

\epostfig{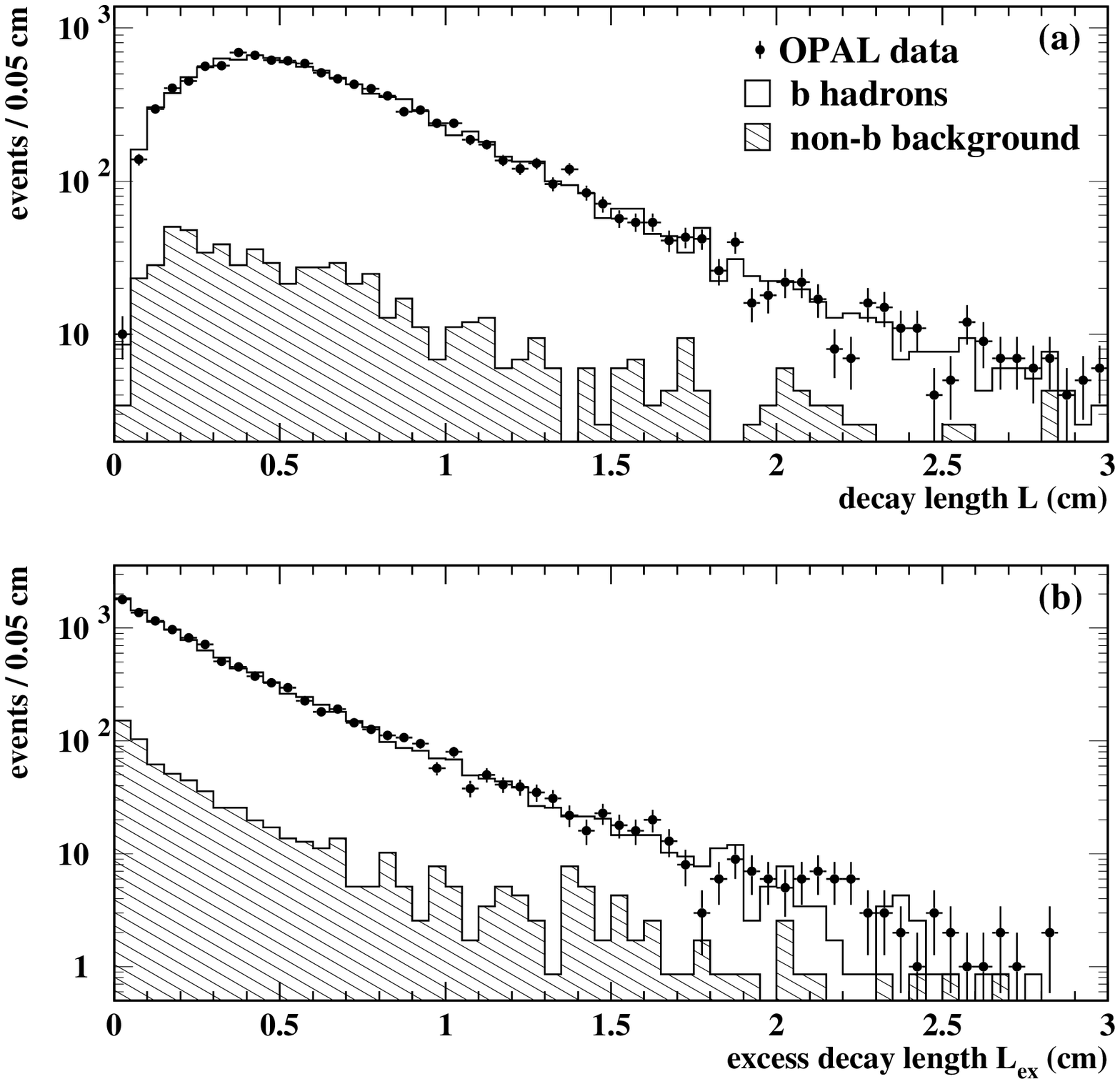}{f:decl}{Distributions of (a) decay length $L$ and (b)
  excess decay length \lex\ in data (points with error bars) and Monte
  Carlo (histogram), for all selected secondary vertices. The expected
  contributions of b hadrons and non-b background are indicated.}

For most selected vertices, the point of failure occurs when one of
the less good arrangements has an improved $\chi^2$, and fails
requirements~1 or~2. Others fail because the $L/\sigma$ separation
becomes too small or the vertex charge error $\sigma_Q$ becomes too
large.

The distribution of excess decay length is approximately a negative 
exponential, with the same slope as the b hadron decay
length distribution before the vertex selection requirements were
imposed. The distribution is exactly an exponential 
provided that the assignment of b hadron decay tracks to the
secondary vertex and fragmentation tracks to the primary vertex is
correct, and that the effects of vertex resolution and the
lifetime of the charm hadron produced
in the b hadron decay can be neglected.

That the distribution is exponential is most easily seen in terms of
the excess decay proper time, obtained from the excess decay length
via the reconstruction of the b hadron energy.
The effect of the charm hadron lifetime will be considered first.
Ignoring resolution effects, the rate of
events $F(t)$ with excess decay time $t$ is given by the convolution
of the lifetime exponentials $e^{-{t_b}/\tau_b}$ and
$e^{-t_c/\tau_c}$ for the decaying b and charm hadrons. Here $t_b$ is
the decay time and $\tau_b$ the lifetime of the b hadron, and
similarly for the charm hadron. 
The convolution is obtained by integrating over 
the excess decay time $t'$ of the b hadron, defined as $t'=t_b-t_0$, where
$t_0$ is the minimum decay time for this event to pass the vertex selection
requirements. Since the introduction of
excess decay time just corresponds to the redefinition of zero time
at $t_b=t_0$ and does not affect the form of the b lifetime
exponential, the distribution $F(t)$ is given by replacing $t_b$ by
$t'$, $t_c$ with $t-t'$, and integrating:
\begin{eqnarray*}
F(t) & \propto & \int_a^b e^{-t'/\tau_b} e^{-(t-t')/\tau_c}\,dt' \\
 & \propto & e^{-t/\tau_c}\ \left[ e^{-t' (1/\tau_b-1/\tau_c)} 
\right]_{t'=a}^{t'=b}
\end{eqnarray*}
where normalisation factors have been neglected.
The upper limit $b$ is simply given by $b=t$, as only b hadrons
decaying with excess time smaller than $t$ can contribute. The lower
limit $a=-t_0$, corresponding to b hadrons decaying at $t_b=0$. 
As long as this minimum b hadron excess decay
time is in magnitude much larger than the 
maximum contributing charm decay time, the lower limit can be approximated
by $a=-\infty$.
The requirements of Section~\ref{ss:lfvtx} select events with a single
secondary vertex well separated from the primary vertex. This  ensures a
long b hadron decay time, and suppresses events with a resolvable
tertiary vertex from a long-lived charm hadron, effectively truncating
the charm decay exponential. These conditions ensure that the 
approximation $a=-\infty$ is valid,
and the integral finally becomes
\[
F(t) \propto e^{-t/\tau_b}\ .
\]
A similar argument holds for the effect
of the finite detector resolution. In this case, the resolution function
replaces the charm hadron lifetime
exponential in the convolution, and the limits become
$a=-\infty$ and $b=\infty$. Since the convolution of an exponential
with any finite function is another exponential with the same decay
constant as the original, the b hadron lifetime distribution is 
again recovered.

However, for a small fraction of events
(mainly those with low \lex ), one or more tracks are mis-assigned from
the primary to the secondary vertex or {\em vice versa\/},
which introduces distortions in the excess decay length and proper
time distributions. Such events
also tend to have some tracks with vertex charge weights 
$w_i\approx 0.5$ ({\em i.e.\/} not clearly assigned to either vertex), 
and so are concentrated in the regions of $Q$ away from 
the peaks at integer values.

The excess decay length was combined with an estimate of the b hadron
energy $E_{\rm b}$ in each event to determine the excess proper time $t$, 
via the relation:
\[
t=\frac{m_{\rm b}\lex}{\sqrt{E^2_{\rm b}-m^2_{\rm b}}}
\]
where $m_{\rm b}$ is the b hadron mass, taken to be that of the
\bplus\ and \bzero, {\em i.e.\/} 5.279\,GeV \cite{pdg98}. The b hadron
energy was estimated using a technique described in \cite{opalbinc}.
First, the energy of the b jet $E_{\rm bjet}$ 
was estimated by treating the event as
a two-body decay of a \zb\ into a b jet of mass $m_{\rm b}$ and
another object making up the rest of the event. Then, the charged and
neutral fragmentation energy $E_{\rm bfrag}$ in the b jet was estimated
using the charged track
weights $w_i$, and the unassociated electromagnetic calorimeter
clusters weighted according to their angle with respect to the jet
axis. Finally the b hadron energy was calculated as
$E_{\rm b}=E_{\rm bjet}-E_{\rm bfrag}$.

The resulting distributions of b hadron energy in data and Monte Carlo
are shown in Figure~\ref{f:energy}(a). The agreement is generally
good, and the small differences around the peak are within the
uncertainties due to the imprecise knowledge of b fragmentation.
In addition to all the vertex
requirements described above, selected M-tag hemispheres were required
to have $E_{\rm b}>20$\,GeV, which suppresses to a negligible level 
events where the M-tagged jet originates from a gluon and not a b quark.
This cut is included in the definition of
the M-tag and is included in the 
event counts and purities of Table~\ref{t:purity}.
The energy resolution in Monte Carlo \bbbar\ events is shown in
Figure~\ref{f:energy}(b). The reconstructed energy has a mean equal
to the true energy, but
has asymmetric tails. However, these tails are small enough not to
have a significant effect on this analysis.

\epostfig{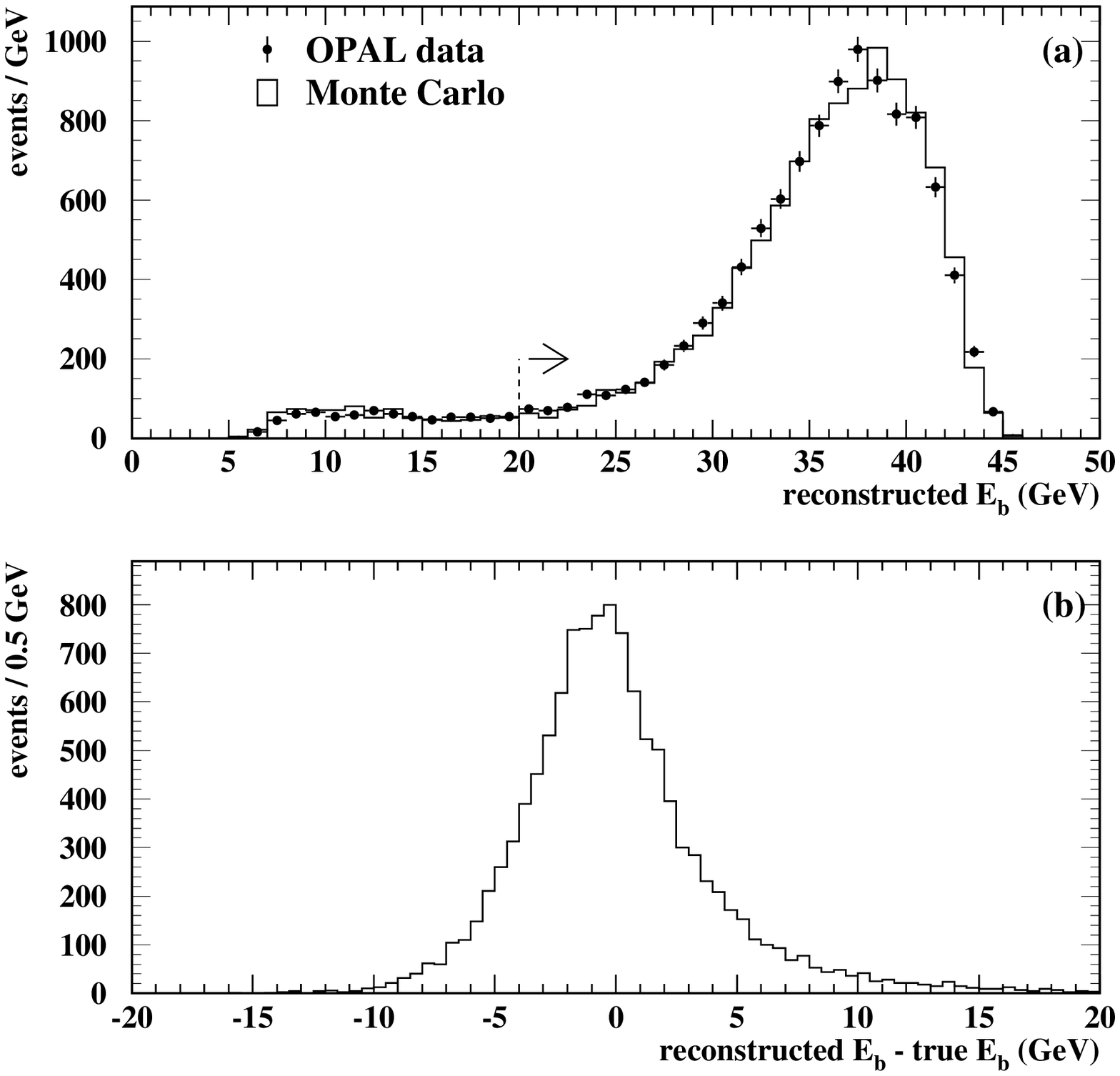}{f:energy}{(a) Reconstructed b hadron energy
  distributions for data (points with error bars) and Monte Carlo
  (histogram). The position of the minimum energy cut is shown by the
  arrow. (b) Resolution of the reconstructed b hadron energy in Monte
  Carlo \bbbar\ events, after the minimum energy cut has been applied.}

\subsection{Fit and results}

The lifetimes of the \bplus\ and \bzero\ mesons were extracted by using
a maximum likelihood fit to the mean excess proper time $t$ as
a function of the modulus of the vertex charge $Q$. 
All events with $0<t<15\rm\,ps$
were used in the fit. The data were divided into ten bins between 
$|Q|=0$ and $|Q|=2$, and all events with $|Q|>2$  put into an
eleventh bin. The mean excess decay time $\mean{t}_j$ was then
calculated for each $|Q|$ bin $j$, and compared to the fit prediction
$\tau_j$. The latter depends on the lifetime $\tau_s$
of each source $s$ (b hadron type or background) 
and the fraction of each source $\fsj$ expected in bin $j$.

The fractions $\fsj$ depend on the vertex charge $Q$. If
the charge tagging were perfect, only charged b hadrons would be
expected at $Q=\pm 1$, and only neutral b hadrons at $Q=0$. However,
there is some cross-contamination, as can be seen in
Figure~\ref{f:qvtx}. As the measurement of the \bplus\ and \bzero\ 
lifetimes depends crucially on the level of this contamination, this was
fitted from the data by exploiting the correlation of the vertex
charge $Q$ with the opposite (T-tag) hemisphere b production flavour
estimate \qt. For example, a b hadron in the T-tag hemisphere implies
a \bqbar\ hadron\footnote{By convention, the b mesons considered as
  particles (\bplus, \bzero, \bs) contain a \bqbar\ antiquark, and the
antiparticles (\bminus, \bzerobar, \bsbar) contain a b quark.
The opposite is true for baryons, so a \bbary\ contains a b quark and
a \bbarybar\ contains a \bqbar\ antiquark.}
(\bminus, \bzerobar, \bsbar\ or \bbary)
in the M-tag hemisphere, giving a correctly reconstructed charge of
$Q=-1$ or $Q=0$. The number of such events reconstructed with $Q=+1$ 
therefore gives information on the number of neutral b hadrons 
incorrectly reconstructed as charged b hadrons, since a true 
\bplus\ (being a b rather than a \bqbar\ hadron) cannot be opposite
another b hadron. In fact, the charged/neutral separation is a
function of the continuous variable $Q$, and is described by a
parameterisation constrained by the above
correlation. These parameters were determined at the same time as the 
source lifetimes $\tau_s$ as additional parameters in the fit.

In more detail, the total likelihood of the event sample was given by:
\begin{equation}
{\cal L}=\ltime \cdot \prod_i \ltag_i \label{e:ltot}
\end{equation}
where $\ltime$ represents the likelihood from the fit to the lifetime as a 
function of vertex charge $Q$ and $\ltag_i$ represents the likelihood
of each event used to determine the
charged/neutral separation. The former is calculated in bins of $Q$,
whilst the latter is determined event by event, and the product is
taken over all events $i$.

The logarithm of the time likelihood $\ltime$ is given by:
\[
\ln\ltime = \sum_j -N_j \left( \frac{\mean{t}_j}{\tau_j}+\ln\tau_j \right)
\]
where the index $j$ runs over the bins of vertex charge $Q$.
The term inside the sum represents the log-likelihood to measure a 
mean decay time $\mean{t}_j$ in a sample of $N_j$ events distributed
according to a negative exponential with lifetime $\tau_j$. Although
the data events are not distributed according to a single exponential,
the differences are sufficiently small that this expression can be used.
The expected true mean decay time in bin $j$ is given by:
\begin{equation}
\tau_j= D(\mean{Q}_j) + \sum_s \fsj \tau_s \label{e:timej}
\end{equation}
where $\fsj$ is the fraction of events from source $s$ in bin $j$, and
$\tau_s$ is the lifetime of source $s$. There are nine sources in
total:
\[
s = \{ \bplus, \bminus, \bzero, \bzerobar, \bs, \bsbar, \bbarybar,
\bbary, \mbox{background} \}
\]
and the lifetimes of particle and antiparticle are assumed to be equal
($\taubp=\taubm$ {\it etc\/}). The background is characterised by a
lifetime \taubg, taken from Monte Carlo.
 
The function $D(\mean{Q}_j)$ in equation~\ref{e:timej} accounts for
the distortions in the excess decay length distribution caused by 
mis-assigned tracks discussed in Section~\ref{ss:tlex}. These can be
seen in Figure~\ref{f:tvq}, which shows the mean excess proper time
$\mean{t}$ as a function of $|Q|$ for both data (points) and Monte
Carlo (solid line).
Events which have one or more mis-assigned tracks tend to have
smaller than average excess proper time, and to be concentrated away
from integer values of $Q$. This reduces the mean excess decay time in these 
regions, and consequently increases the mean time close to integer
values of $Q$. This effect, seen clearly in both data and Monte Carlo,
is parameterised by the periodic correction function $D$, which has the form:
\begin{equation}
D(q)=d(q-1/2)^2-d/8 \ \ \{ \mbox{for $0<q<1$} \} \label{e:gdist}
\end{equation}
where $q=(Q-{\rm int}(Q))$ is the non-integer part of $Q$. 
The parameter $d$
characterises the amplitude of the distortion, and is left as a free
parameter in the fit. The functional form of this correction 
was chosen by studying the effect in Monte Carlo.

\epostfig{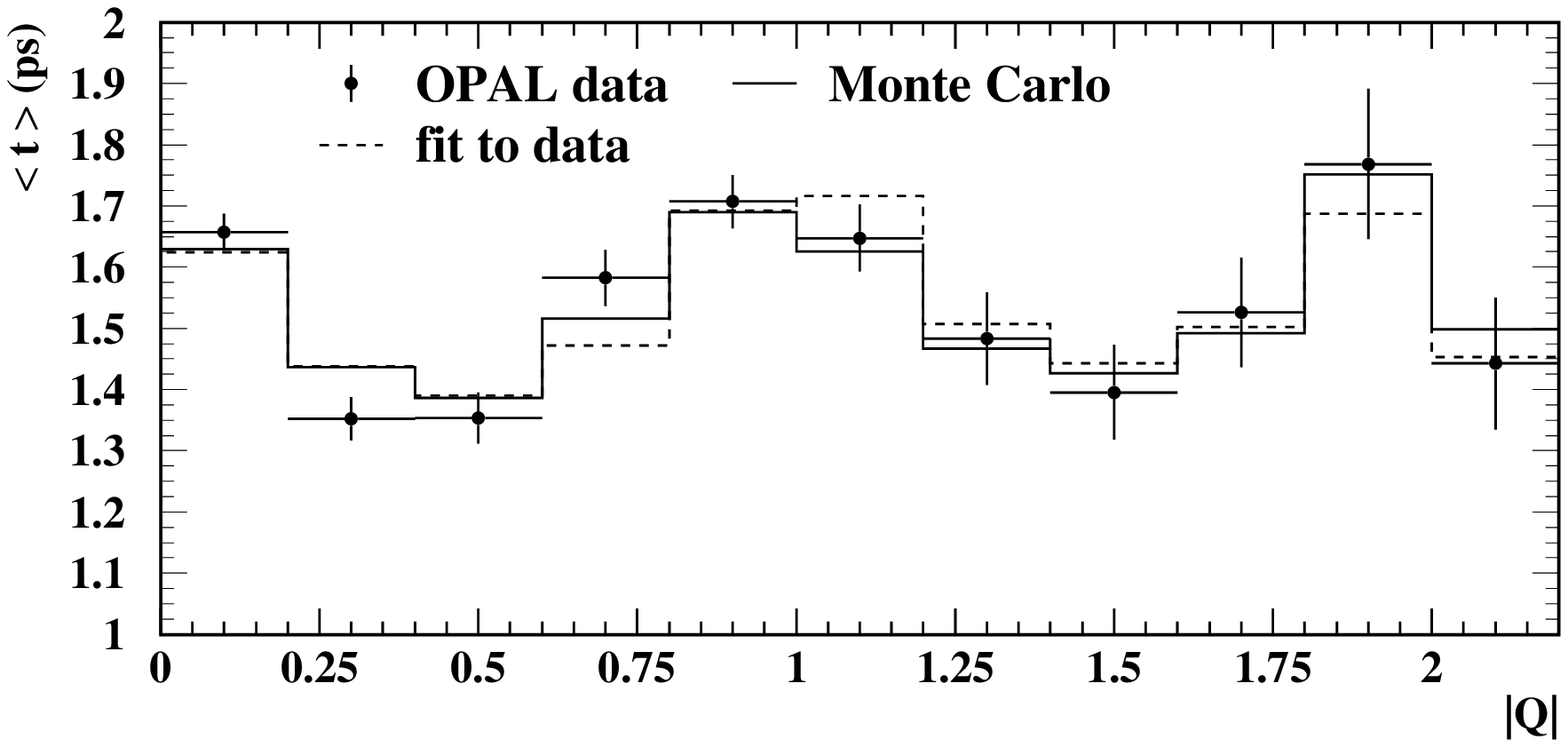}{f:tvq}{The distribution of mean excess decay time
$\mean{t}$ as a function of the modulus of the
vertex charge $|Q|$. The data are
shown by the points with error bars, and the Monte Carlo (reweighted to
the same \bplus\ and \bzero\ lifetimes as measured in the data) 
distribution is
shown by the solid line. The  prediction of the fit is shown by
the dashed line. The bin between $|Q|=2$ and $|Q|=2.2$ contains all events
with $|Q|>2$.}

The second part of the overall likelihood in equation~\ref{e:ltot} is
the tag likelihood $\ltag$. It is given for each event $i$ by:
\begin{equation}
\ltag_i = \sum_s \psq (Q_i)\, \pst (\qt_i) \label{e:ltag} 
\end{equation}
The function $\psq(Q)$ gives the probability of each source $s$ as a 
function of the b hadron vertex charge $Q$. For the \bplus\, this 
function is given by:
\[
P^Q_{\rm B^+}(Q) = \left\{ \begin{array}{ll}
(1-\fbg)c_3 & \mbox{for $Q\le -1$} \\ 
(1-\fbg)(c_0+(c_0-c_3)Q) & \mbox{for $-1<Q\le 0$} \\
(1-\fbg)(c_0+(c_1-c_0-c_2)Q+c_2Q^2) & \mbox{for $0<Q<1$} \\
(1-\fbg)c_1 & \mbox{for $Q\ge 1$} 
\end{array} \right. 
\]
where \fbg\ is the fraction of non-\bbbar\ background, $c_0$ represents
the fraction of signal \bplus\ events at $Q=0$, and $c_1$ the fraction
of signal \bplus\ events at $Q=1$. A quadratic interpolation with coefficient
$c_2$ is used in the range $0<Q<1$, and the fraction of \bplus\ is 
constant for $Q\ge 1$. 
A small fraction $c_3$ of \bplus\ events has $Q\le -1$, 
and a linear interpolation is used for the fraction of \bplus\ between
$-1$ and $0$. This functional form is illustrated in
Figure~\ref{f:vqvfit} and is seen to give a reasonable description
of the \bplus\ fraction in Monte Carlo. Large variations in the
parameters $c_0$ and $c_3$ are considered when assessing the
systematic errors.

\epostfig{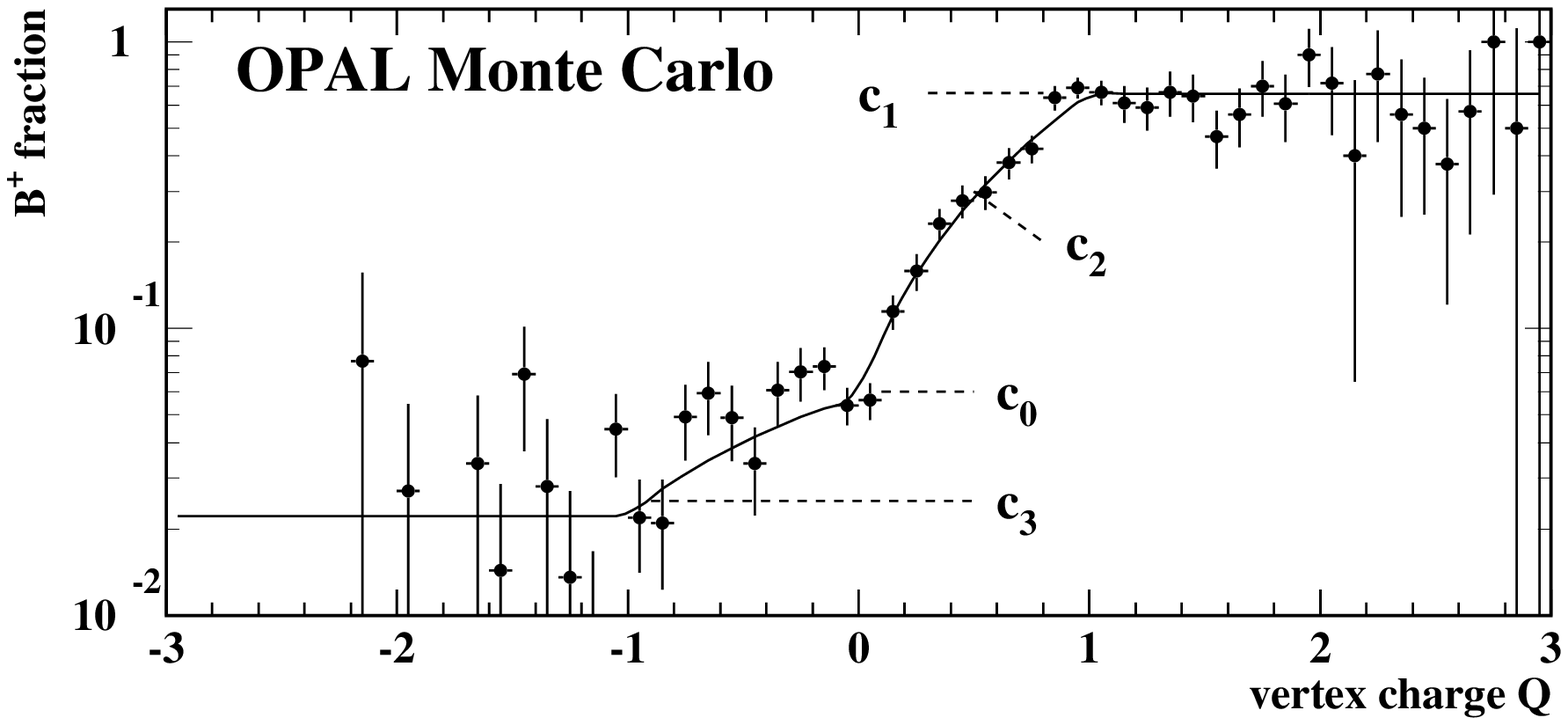}{f:vqvfit}{Fraction of \bplus\ vertices as a
  function of vertex charge $Q$ in Monte Carlo \bbbar\ events (points
  with error bars). The parameterisation is shown by the solid line,
  and the levels $c_0$, $c_1$ and $c_3$ by the dotted lines. The
  quadratic coefficient $c_2$ is used between $Q=0$ and $Q=1$. Note
  that the non-\bbbar\ background fraction is not included, and the
  \bplus\ fraction is plotted using a logarithmic scale.}

The fraction of \bminus\ is given by charge symmetry: 
$P^Q_{\rm B^-}(Q)=P^Q_{\rm B^+}(-Q)$. The remaining signal fraction is 
neutral b hadrons, so the fractions of \bzero\ and \bzerobar\ are given by:
\begin{equation}
P^Q_{\rm B^0}(Q) = P^Q_{\rm \bar{B}^0}(Q) = 
(1-\fzbs-\fzbbary)(1-\fbg-P^Q_{\rm B^+}(Q)-P^Q_{\rm B^-}(Q)) \label{e:psqn}
\end{equation}
where $\fzbs$ and $\fzbbary$ are the fractions of \bs\ and \bbary\ in
the neutral b hadron sample. The $\psq$ values for \bs\ and \bbary\
are given in an analogous way to equation~\ref{e:psqn} but with the factor
$(1-\fzbs-\fzbbary)$ replaced by $\fzbs$ or $\fzbbary$. Finally
the background function $P^Q_{\rm bg}(Q)=\fbg$, {\em i.e.\/} the background
fraction is a constant independent of $Q$, as found in Monte Carlo.

These functions together describe the source fractions as a function of $Q$ in
terms of four parameters: $c_1$ and  $c_2$, which are left free in the fit;
and $c_0$ and $c_3$, which are  input from Monte Carlo. The values of 
$c_1$ and $c_2$ are constrained by the correlation between
the type of b hadron in the M-tag hemisphere and the production
flavour (b or \bqbar) of the other b hadron in the T-tag hemisphere.
The contamination of neutral b hadrons at $Q\approx\pm 1$ 
(which is given by $1-c_1-c_3$) is thus determined
almost entirely from the data, whilst the contamination of charged b 
hadrons at $Q\approx 0$ (given by $c_0$) has to be taken from Monte
Carlo, since the vertex
charge provides no distinguishing power between neutral b and \bqbar\
hadrons.

The function $\pst(\qt)$ in equation~\ref{e:ltag} gives the probability 
for each source to 
be tagged by the opposite hemisphere flavour tag of value \qt:
\[
\pst (\qt) = \left\{ \begin{array}{ll}
(1-\qt)/2 & \mbox{for $s=\bplus, \bzero, \bs, \bbarybar$} \\
(1+\qt)/2 & \mbox{for $s=\bminus, \bzerobar, \bsbar, \bbary$} \\
1 & \mbox{for $s=$background}
\end{array} \right.
\]
Finally, the fractions $\fsj$ of each source contributing in each bin $j$
of the time likelihood (equation~\ref{e:timej}) are given by the average of 
the source fractions $f^i_s$ for each of the events $i$ in bin $j$,
where:
\[
f^i_s=\frac{\psq(Q_i) \pst(\qt_i)}{\sum_{s'} P^Q_{s'}(Q_i) P^T_{s'}(\qt_i)}
\]

In total, 5~parameters were left free in the fit: the \bplus\ and
\bzero\ lifetimes \taubp\ and \taubz, the charge separation parameters
$c_1$ and $c_2$, and the distortion parameter $d$. 
The \bs\ and \bbary\ lifetimes were taken to be $1.54\pm 0.07\rm\,ps$
and $1.22\pm 0.05\rm\,ps$ \cite{pdg98}. The background lifetime was taken to
be that in the Monte Carlo, $\taubg=2.1\rm\,ps$, and the background
fraction was measured to be $\fbg=0.052\pm 0.015$ 
(see Table~\ref{t:purity}). The remaining parameters $\fzbs$, $\fzbbary$,
$c_0$ and $c_3$ were taken from the Monte Carlo. Uncertainties on
all these parameters are considered when evaluating the systematic errors.

The results of the fit for the \bplus\ and \bzero\ lifetimes are:
\begin{eqnarray*}
\taubp & = & \tpval \pm \tpstat \rm\,ps \\
\taubz & = & \tzval \pm \tzstat \rm\,ps 
\end{eqnarray*}
where the errors are statistical only. The correlation coefficient 
between the two measured lifetimes is $-0.53$, and their ratio is
$\taubr=\trval\pm\trstat$. The other fitted parameters were measured
to be $c_1=0.84\pm 0.03$, $c_2=0.10\pm 0.18$ and $d=1.76\pm 0.21$\,ps.

The distribution of mean excess decay time $\mean{t}$ as a function of
$|Q|$ is shown in Figure~\ref{f:tvq}, for data and for Monte Carlo
simulation
reweighted to the measured \bplus\ and \bzero\ lifetimes. The Monte
Carlo sample is 13 times larger than the data sample, and the
$\chi^2$ between data and Monte Carlo distributions is 9.4 for 8
degrees of freedom. The results of the fit to the data (the values of
$\tau_j$) are shown by the dashed line, and the fit follows the data 
reasonably well. The distribution of data excess
proper time in various regions of $|Q|$ is shown in
Figure~\ref{f:propt}, again together with the reweighted Monte Carlo 
simulation. The deviations from a pure exponential away from integer values
of $Q$ can clearly be seen in both data and Monte Carlo.

\epostfig{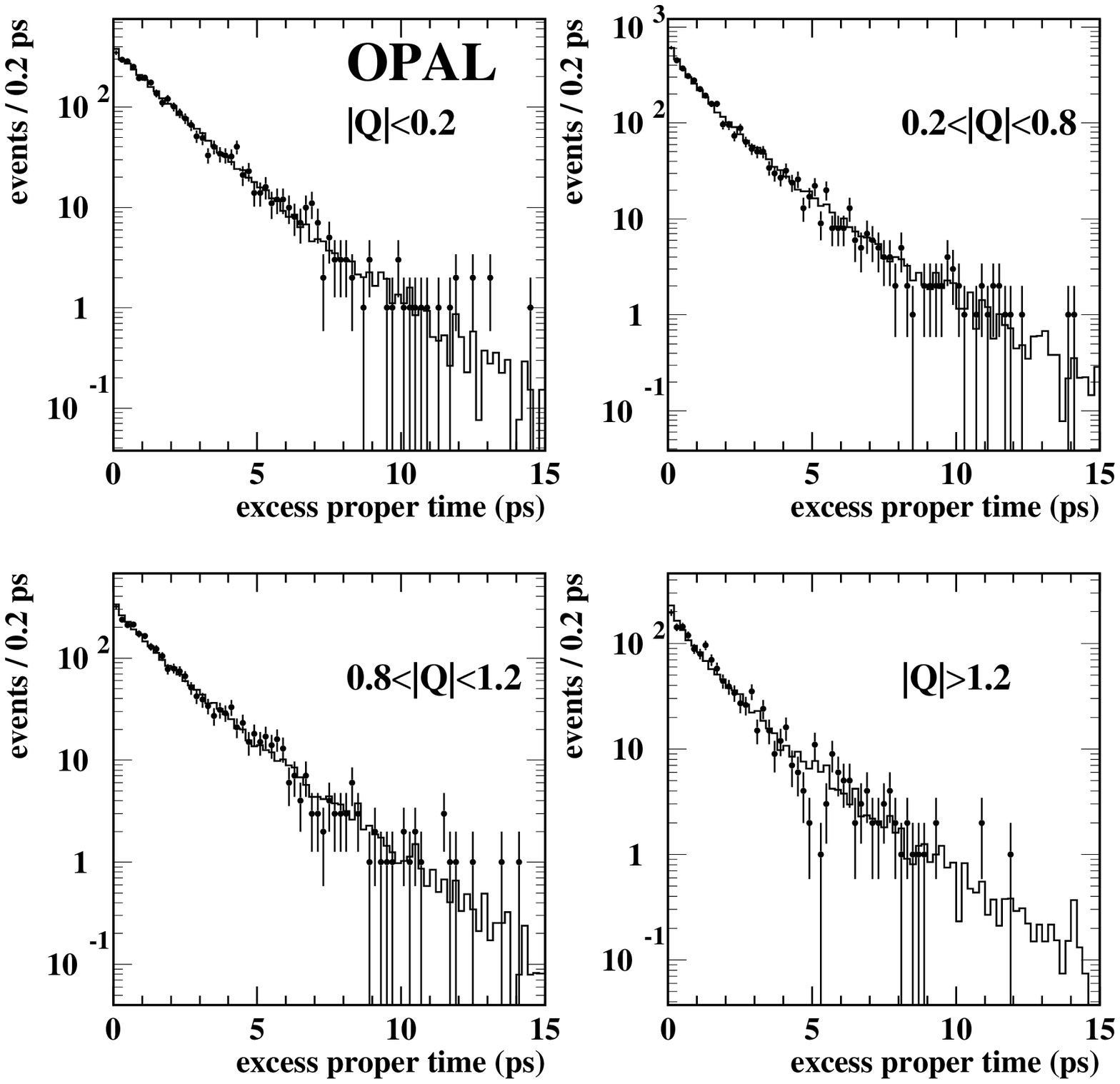}{f:propt}{Distributions of excess proper time in
  four regions of absolute vertex charge $|Q|$, 
  for data (points with error bars)
  and Monte Carlo simulation (histogram). The Monte Carlo \bplus\ and
\bzero\ lifetimes have been reweighted to the same values as measured in
  the data.}

\subsection{Systematic errors}\label{ss:lfsyst}

The systematic errors on the \bplus\ and \bzero\ lifetime measurements
and their ratio are summarised in Table~\ref{t:syst}, and 
discussed in more detail below. In most cases, the systematic errors
are evaluated by repeating the fit in either data or Monte Carlo, 
varying the appropriate parameter.

\begin{table}
\centering

\begin{tabular}{l|ccc}\hline\hline
Source & $\Delta\taubp$\,(ps) & $\Delta\taubz$\,(ps) & $\Delta\taubr$ \\ \hline
\bs\ lifetime           & 0.000 & 0.026 & 0.019 \\
b baryon lifetime       & 0.000 & 0.011 & 0.008 \\
$\bs/\bzero$ fraction   & 0.000 & 0.001 & 0.001 \\
$\bbary/\bzero$ fraction & 0.000 & 0.025 & 0.018 \\
Background lifetime     & 0.005 & 0.009 & 0.003 \\
Background fraction     & 0.010 & 0.015 & 0.004 \\
Fit procedure           & 0.014 & 0.023 & 0.022 \\
Charge separation       & 0.005 & 0.019 & 0.015 \\
b fragmentation         & 0.008 & 0.009 & 0.000 \\
Detector resolution     & 0.008 & 0.007 & 0.009 \\
Silicon alignment       & 0.009 & 0.011 & 0.012 \\
\hline
Total                   & \tpsyst & \tzsyst & \trsyst \\
\hline
\end{tabular}
\caption{\label{t:syst}Systematic errors on the measured values of 
\taubp, \taubz\ and \taubr. The values for \taubr\ take into account
correlations between \taubp\ and \taubz.}
\end{table}

\begin{description}
\item[$\rm\bf B_s$ and $\rm\bf\Lambda_b$ contamination:] 
  The neutral b hadron sample
  consists not only of \bzero\ mesons but an irreducible contribution 
  from \bs\ and \bbary. Uncertainties in both the lifetimes
  ($\tau_{\rm B_s}$ and $\tau_{\Lambda_b}$) and
  sizes ($\fzbs$ and $\fzbbary$) of these backgrounds affect 
  the \bzero\ lifetime measurement, the \bs\ lifetime being the
  largest single source of uncertainty. The efficiencies for \bzero, \bs\
  and \bbary\ to pass the vertex selection requirements are similar,
  but somewhat sensitive to their average charged decay
  multiplicities. However, by far the largest uncertainty in the level
  of contamination comes from
  the knowledge of the \bs\ and \bbary\ production fractions in 
  $\zb\rightarrow\bbbar$ events, taken to be 
  $f({\rm b}\rightarrow\bs)=10.5^{+1.8}_{-1.7}\,\%$ and 
  $f({\rm b}\rightarrow\bbary)=10.1^{+3.9}_{-3.1}\,\%$ \cite{pdg98}.

\item[Non-$\rm\bf b\overline{b}$ background contamination:] 
  The effective mean decay time of
  the non-\bbbar\ contamination in the selected sample is 2.1\,ps in
  Monte Carlo. This
  has two distinct components---a contribution from mis-measured
  tracks and strange particle decays 
  causing fake b decay vertices with long decay lengths, 
  and a contribution
  from the decay of charm hadrons which have genuine lifetime. Charm events
  have both contributions, giving an effective lifetime of about
  1.8\,ps. Only the former contribution is present in uds events,
  which have an effective lifetime of about 2.3\,ps.

To study the modelling of the background lifetime, two
control samples were used, generated by modifying the selection for
the T-tag. Requiring the vertex tag variable $B$ in the T-tag 
to be between 0 and 1.2 gave a sample consisting of 9\,\% uds, 18\,\%
\ccbar\ and 72\,\% \bbbar\ events, and requiring the T-tag hemisphere
to fail the vertex tag pre-selection \cite{opalrb} gave a sample
of 41\,\% uds, 23\,\% \ccbar\ and 35\,\% \bbbar\ events. The effective
lifetimes measured in the M-tag hemispheres for these samples were
found to agree in data and Monte Carlo to better than  
0.05\,ps. From this, an upper limit due to the mis-modelling
of the background in the primary b-tagged sample was estimated as
0.1\,ps, and used to assess the systematic error on the \bplus\ and
\bzero\ lifetimes.

\item[Fit procedure:] The entire fitting procedure, including the
  derivation of the excess decay lengths and the
  correction for the distortion given in equation~\ref{e:gdist}, was
  tested on a fully simulated Monte Carlo sample 13 times larger than
  the data sample. In this sample, where the \bplus\ and \bzero\
  lifetimes were both 1.6\,ps, the fit gave the results
  $\taubp=1.588\pm 0.014$\,ps and $\taubz=1.623\pm 0.017$\,ps. The
  larger of the deviations of these results from the true values or 
  the statistical errors were taken as
  systematic errors due to the fitting procedure. As can be seen from 
  Figures~\ref{f:tvq} and~\ref{f:propt}, the Monte Carlo provides a
  good description of the time distributions in the data.

Additional Monte Carlo studies were performed to check the correctness
of the fit procedure and errors. To verify the errors returned by
the fit, it was performed on many Monte Carlo subsamples, and the
distribution of fitted lifetimes studied. The Monte Carlo was
also reweighted to change the \bplus\ and \bzero\ lifetimes one at a time.
In each case, the fit correctly recovered the modified lifetime,
whilst returning a stable result for the lifetime of the unmodified b hadron.

As a cross check, the fit to the data was repeated with the
parameterisation of the distortion given in equation~\ref{e:gdist}
replaced with the distortions measured in the large Monte Carlo
sample. Consistent results were obtained, with the values of 
\taubp\ and \taubz\ changing by 0.017\,ps and $-0.013$\,ps respectively.

\item[Charge separation:] The fraction of neutral b hadron contamination
  at $|Q|\approx 1$ ({\em i.e.\/} $1-c_1-c_3$) was
  fitted from the data. However, the corresponding contamination of
  charged b hadrons at $Q\approx 0$ ({\em i.e.\/} $c_0$), and the
  fraction of wrong sign charged b hadrons $c_3$ were
  taken from the Monte Carlo, and varied by $\pm 50\,\%$ to assess the
  systematic error (see Figure~\ref{f:vqvfit}).
  This variation is larger than the difference seen
  between data and Monte Carlo in the neutral contamination of the
  charged b sample, and also larger than the variation resulting from 
  changing the b hadron charged decay multiplicities within their 
  experimental uncertainties \cite{hfew}.

  The fit to determine the charge separation depends on the T-tag
  hemisphere production flavour tag \qt, which is subject to mis-tag
  and jet and vertex charge offset subtraction uncertainties, as discussed in
  Section~\ref{s:bprod}. The effect of these uncertainties on the
  b hadron lifetime measurement is negligible.

\item[b fragmentation:] The effect of uncertainties in the average b
  hadron energy $\meanxe=E_{\rm b}/{E_{\rm beam}}$ was assessed in
  Monte Carlo by reweighting so as to vary $\meanxe$ in the range
  $0.702\pm 0.008$ \cite{hfew}, and repeating the lifetime fit.
  The effect on the lifetimes is small since the b hadron
  energy is estimated event by event.

\item[Detector resolution and alignment:]
The error due to uncertainty in the tracking detector resolution was 
assessed in Monte Carlo
by applying a global $10\,\%$ degradation to the resolution of
all tracks, independently in the $r$-$\phi$ and $r$-$z$ planes, as in 
\cite{opalrb}. The lifetime measurements are also sensitive to the effective
radial positions of the silicon detectors (both the positions of the detectors
themselves and the positions of the charge collection regions within
them). These are known to a
precision of $\pm 20\rm \,\mu m$ from studies of cosmic ray events
\cite{opalrb}. The resulting uncertainty was calculated by applying 
$20\rm\,\mu m$ radial shifts to one or both silicon layers in Monte Carlo and
repeating the lifetime fits.

\end{description}

The total systematic errors amount to $\pm \tpsyst$\,ps on the \bplus\
and $\pm \tzsyst$\,ps
on the \bzero\ lifetimes, and $\pm \trsyst$ on their ratio,
where correlated systematic errors have been taken into account.

\section{CP(T) violation analysis}\label{s:cpt}

In the Standard Model framework for describing CP violation, according
to the formalism given in equation~\ref{e:eigen}, an
asymmetry is predicted in the time dependent inclusive decay rates of
\bzero\ and \bzerobar\ mesons:
\[
A(t') \equiv \frac{
 \Gamma(\bzero\rightarrow\mbox{anything})-
 \Gamma(\bzerobar\rightarrow\mbox{anything})}
{\Gamma(\bzero\rightarrow\mbox{anything})+
\Gamma(\bzerobar\rightarrow\mbox{anything})}
\]
For a totally unbiased selection of \bzero\ and \bzerobar\  mesons,
the form of this asymmetry as a function of true proper decay time 
$t'$ is given by:
\begin{equation}
A(t')= \acp \left\{ \frac{\dmd\taubz}{2} \sin (\dmd t') - 
\sin^2\left(\frac{\dmd t'}{2}\right) \right\} \label{e:asymt}
\end{equation}
where \acp\ is the CP-violating observable, \dmd\ is the \bzero\
oscillation frequency and \taubz\ the \bzero\
lifetime\cite{cpinc,dun2}. The parameter
\acp\ is related to the CP-violation parameter \epsb\ by:
\begin{equation}
\repsb = \frac{\acp}{4} \label{e:repsb}
\end{equation}
using the convention that $|\epsb|$ is small. 
Searching for an asymmetry of the form given in 
equation~\ref{e:asymt} therefore provides a method of probing the
value of \epsb. In the Standard Model, \repsb\ is expected to be
around $10^{-3}$ \cite{cpinc}, but it could be up to an order of 
magnitude larger in superweak models \cite{superw}.

In this analysis, the T-tag is used to tag \bbbar\ events, and an
inclusive reconstruction algorithm is used to reconstruct the b hadron
decay time $t$ in the M-tag hemisphere. The production flavour (b or
\bqbar) of this M-tagged b hadron is given by the combined tagging variable
\qe\ (see section~\ref{s:bprod}), using information from both
hemispheres of the event. No attempt is made to separate \bzero\ and 
\bzerobar\ decays from other b hadron decays in the M-tag hemisphere.
A similar asymmetry to that given in equation~\ref{e:asymt} is
expected for \bs\ mesons, with \dmd\ replaced with the \bs\
oscillation frequency \dms, but it is expected to be at least an order
of magnitude smaller and is neglected. No asymmetry is expected for \bplus\ and
\bbary. These other b hadrons therefore dilute the expected \bzero\
asymmetry, but do not change its form. Hence no attempt is made to remove
\bplus\ decays by requiring well reconstructed neutral vertices, as this would
not greatly increase the sensitivity to CP violation in the \bzero\ decays,
and would lead to a large loss in reconstruction efficiency.

Although the time dependent rates of \bzero\ and \bzerobar\ decay are
predicted to be different, this does not violate the CPT invariance
 prediction of
equal total decay rates. The lifetimes of \bzero\ and \bzerobar\ are thus
still expected to be equal. If CPT violation were present, the
lifetimes of b and \bqbar\ hadrons could be different:
\begin{eqnarray}\label{e:deltab}
\taub & = & \left\{1+\frac{1}{2}\deltabbg\right\} \tauav \\
\taubbar & = & \left\{1-\frac{1}{2}\deltabbg\right\} \tauav \nonumber
\end{eqnarray}
where \tauav\ is the average and \deltab\ the fractional 
difference in lifetimes.
The value of \deltab\ is determined by measuring the lifetimes of b and
\bqbar\ hadrons, using the same time 
reconstruction algorithm as for the CP-violation analysis.

\subsection{b hadron reconstruction} \label{ss:cpvtx}

The b hadron decay length was reconstructed using a secondary vertex finding 
algorithm similar to those employed in \cite{opalbstar} and \cite{opalbinc}.
In the highest energy jet in the M-tag hemisphere, the two tracks with the 
largest impact parameter with respect to the primary vertex were used
to form a `seed' vertex. All tracks consistent with this seed vertex
were then added to it using an iterative procedure. Only tracks satisfying the
quality requirements detailed in Section~\ref{ss:lfvtx} were considered for 
inclusion in the secondary vertex.

The resulting vertex was required to have at least three tracks, and the
invariant mass (assuming all tracks to be pions) was required to 
be at least 0.8\,GeV. The transverse miss distance (the projection of
the vector between primary and secondary vertices onto a plane
orthogonal to the total momentum vector of the tracks assigned to the
secondary vertex) divided by its error was required to be less than~3
\cite{opalbinc}. These requirements help to eliminate badly reconstructed and
fake secondary vertices. The decay length $L$ between primary and secondary
vertices was then calculated, using the jet axis direction as a constraint,
as in \cite{opalrb}. An acceptable secondary vertex was found in approximately
70\,\% of hemispheres, for both \bbbar\ and background (\ccbar\ and light 
quark) events.

The b hadron energy was estimated as described in section~\ref{ss:tlex}, 
calculating the charged fragmentation energy using weights $w_i$ tuned
with this alternative vertex finder. The estimate $t$ of the b hadron decay
proper time was then calculated as before:
\[
t=\frac{m_{\rm b}L}{\sqrt{E^2_{\rm b}-m^2_{\rm b}}} \ .
\]
In Monte Carlo, the resolution of this estimate can be described 
by the sum of two Gaussians. For the 1993--95 data, the narrower Gaussian
has an RMS width of 0.33\,ps and contains 65\,\% of the events, and the wider 
Gaussian has a width of 1.3\,ps. For the 1991--92 data, where the vertex
reconstruction was done in the $r$-$\phi$ plane only, the narrow Gaussian
has a width of 0.33\,ps and contains 64\,\% of events, and the wider 
Gaussian has a width of 1.4\,ps. Since the resolutions are similar,
the two data samples were combined and a single resolution function
used for the entire sample. These resolutions are an average over all
true decay proper times $t'$, and significant non-Gaussian effects can be 
seen in small slices of $t'$, as in \cite{opalbinc}. These effects are caused
by the presence of tracks from the primary vertex, and make a full description
of the resolution as a function of $t'$ very complicated. However, they are
not important for the analysis described here, which does not rely heavily on
an accurate description of the decay time resolution. 

\subsection{Fit and results for $\rm\bf\epsilon_B$}\label{ss:cpres}

The CP-violating parameter \acp\ was extracted using a $\chi^2$ fit
to the observed time dependent asymmetry $A(t)$ in 34 bins of 
reconstructed time in the range $-2$ to 15\,ps. Within each time bin
$i$, the observed asymmetry was calculated in ten bins $j$ of $|\qe|$ 
($0<|\qe|<1$) to make best use of the tagging information in each
event. These ten estimates of the asymmetry were averaged with
appropriate weights to calculate the overall observed asymmetry 
\aiobs\ in each time bin $i$. The observed asymmetry was compared
with the expected asymmetry $A^{\rm true}_i$ calculated for a given
\acp, taking into account the time resolution and dilution from 
non-\bzero\ decays in the event sample.

The corrected observed asymmetry in bin $i$ of reconstructed time
$t$ and bin $j$ of $|\qe|$ is given by:
\[
A^{\rm obs}_{ij} = \frac{N^{\rm b}_{ij}-N^{\rm\bar{b}}_{ij}}{\mean{|\qe|}_{ij}
(N^{\rm b}_{ij}+N^{\rm\bar{b}}_{ij})}
\]
and the error $\sigma_{A^{\rm obs}_{ij}}$ is given by:
\[
\sigma_{A^{\rm obs}_{ij}}=
\frac{1-(\mean{|\qe|}_{ij}\,A^{\rm obs}_{ij})^2}
{2 \mean{|\qe|}_{ij}}\ 
\sqrt{\frac{N^{\rm b}_{ij}+N^{\rm\bar{b}}_{ij}}
{N^{\rm b}_{ij}N^{\rm\bar{b}}_{ij}}}
\]
where $N^{\rm b}_{ij}$ ($N^{\rm\bar{b}}_{ij}$) is the number of
events with $\qe>0$ ($\qe<0$). The factor $1/{\mean{|\qe|}_{ij}}$
corrects for the tagging dilution (mis-tagging), which reduces the observed
asymmetry for imperfectly tagged events. The ten estimates 
$A^{\rm obs}_{ij}$ were then averaged, weighting according to 
$(\sigma_{A^{\rm obs}_{ij}})^{-2}$ to derive the corrected observed asymmetry
\aiobs.

The true asymmetry as a function of reconstructed time $t$ is given
by:
\[
A^{\rm true}(t)=(1-\fbg)\fbzero \left\{ \frac{
\int_0^\infty P(t) A(t') R(t-t') dt'}
{\int_0^\infty P(t) R(t-t') dt'} \right\}
\]
where \fbg\ is the fraction of non-\bbbar\ events in the sample, 
\fbzero\ is the fraction of \bzero\ in the \bbbar\ part of the sample,
$P(t')$ is the lifetime exponential $P(t')=(1/\taubz) e^{-t'/\taubz}$,
$A(t')$ is the asymmetry as a function of true proper time $t'$ given in
equation~\ref{e:asymt}, and $R(t-t')$ is the time resolution function.
$R(t')$ is the sum of two Gaussians, with parameters given in
section~\ref{ss:cpvtx}. The expected asymmetry in time bin $i$ was
calculated from the mean reconstructed decay time of all events in 
bin $i$: $A^{\rm true}_i=A^{\rm true}(\mean{t}_i)$.

A binned $\chi^2$ fit was performed to the distribution of
asymmetry as a function of reconstructed time. Both 1991--92 and 1993--95
data samples were used together, setting the background parameter 
to the average impurity of the two samples,  $\fbg=13.7\pm
3.0\,\%$ (see Table~\ref{t:purity}). The values of \taubz\ and \dmd\
were taken to be $\taubz=1.56\pm 0.04$\,ps and 
$\dmd=0.464\pm 0.018\rm\,ps^{-1}$ \cite{pdg98}. 
The parameter \fbzero\ was taken from the Monte Carlo to
be $\fbzero=0.41$.

The observed asymmetry for the \ntmtot\ data events
is shown in Figure~\ref{f:cpasym}, together with the result of the 
one parameter fit to \acp. The fit result is
\[
\acp = \acpval \pm \acpstat
\]
where the error is statistical only.

\epostfig{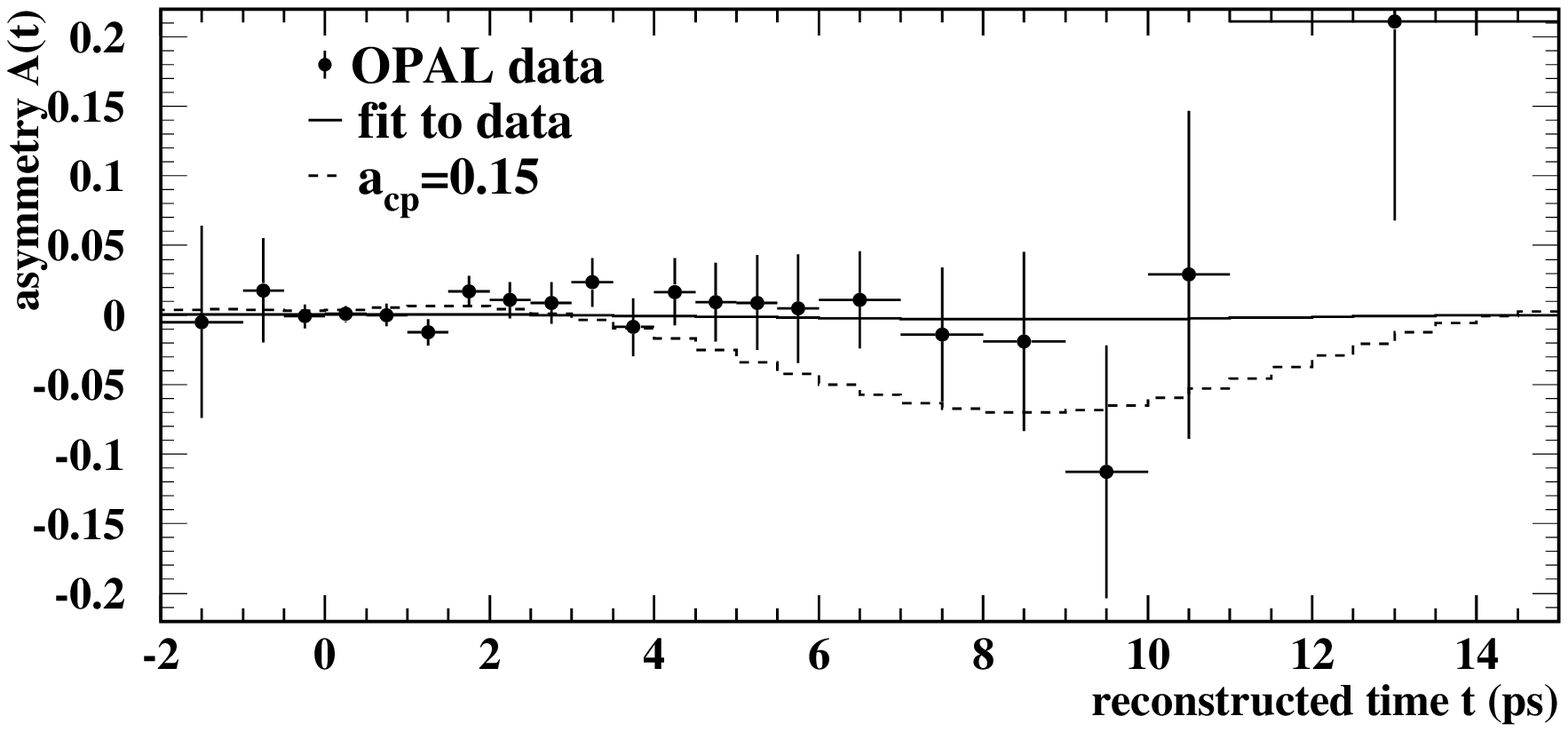}{f:cpasym}{Asymmetry of tagged b and \bqbar\
  hadrons as a function of reconstructed time $t$ in data
  (points), and fit (solid line). The expected asymmetry for
  $\acp=0.15$ is shown by the dotted line.}

The form of the asymmetry given in equation~\ref{e:asymt} assumes that
the reconstruction efficiencies for all decay modes are equal. In
particular, if the efficiency to reconstruct \bzero\ decays to final
states containing no charm hadron, one charm hadron and two charm
hadrons are different, then additional asymmetries may be seen
\cite{dun2}. This is because the semi-inclusive \bzero\ decays to
final states containing different numbers of charm hadrons may exhibit
larger CP-violating effects of different signs which largely cancel
out in the inclusive decay. In this case, the expected asymmetry takes
the form
\begin{equation}
A(t')= \ccp \sin (\dmd t') - 
\acp \sin^2\left(\frac{\dmd t'}{2}\right) \label{e:asym2}
\end{equation}
where \acp\ is still related to \repsb\ as in equation~\ref{e:repsb},
and \ccp\ is an additional CP-violating parameter
\cite{dun2}. A second fit to the form shown in  
equation~\ref{e:asym2}, allowing the values of both \acp\ and \ccp\ to
vary, gave the results
\begin{eqnarray*}
\acp & = & \acptval \pm \acptstat \\
\ccp & = & \ccpval \pm \ccpstat 
\end{eqnarray*}
where again the errors are statistical only, and the correlation 
coefficient between the two parameters is $0.72$. No evidence is seen for a
significant $\sin (\dmd t')$ term, and the value of \acp\ shifts by only
$-0.003$ (5\,\% of the statistical error) 
with respect to the one parameter fit, showing that the effects of
efficiency differences are not significant.

The systematic errors on the values of \acp\ and \ccp\ (from the one
and two parameter fits respectively) are all small compared to the
statistical errors. They are shown in
Table~\ref{t:cpsyst}, and discussed in more detail below.

\begin{table}
\centering

\begin{tabular}{l|cc}\hline\hline
Source & $\Delta\acp$ & $\Delta\ccp$ \\ \hline
\bzero\ lifetime & 0.002 & 0.000 \\
\dmd\ value & 0.001 & 0.001 \\
\bzero\ fraction & 0.002 & 0.002 \\
Flavour tagging offsets & 0.003 & 0.013 \\
Flavour tagging mis-tag  & 0.009 & 0.005 \\
b fragmentation & 0.008 & 0.006 \\
Time resolution & 0.002 & 0.000 \\
\hline
Total & \acpsyst & \ccpsyst \\
\hline
\end{tabular}
\caption{\label{t:cpsyst} Systematic errors on the measured values of
  \acp\ (from the one parameter fit) and \ccp\ (from the two parameter
  fit).}
\end{table}

\begin{description}
\item[Physics input parameters:] The values of the \bzero\ lifetime
  \taubz\ and oscillation frequency \dmd\ were taken from \cite{pdg98}
  and varied within the quoted errors.

\item[$\rm\bf B^0$ fraction:] The fraction of \bzero\ events in the data
  sample depends on the production fractions of \bs\ and \bbary\ in
  \bbbar\ events \cite{pdg98} and on the fraction of non-\bbbar\
  background, determined to be $13.7\pm 3.0\,\%$ from the data.

\item[Flavour tagging:] The offsets in the jet and vertex charges 
used for tagging the \bzero\ production flavour were measured directly in 
the data, as described in Section~\ref{s:bprod}. The uncertainties in these
offsets contribute to the systematic errors on \acp\ and \ccp. 

If a time dependent CP-violating effect were present, it could change the 
value of the offset measured for the T-tag vertex charge
\qvtx, as the vertex tagged sample (being biased towards long
\bzero\ decay times) would contain an unequal mixture of \bzero\ and 
\bzerobar\, which do not have equal vertex charge offsets. With the 
offsets measured in Monte Carlo and a CP-violating effect of $\acp=0.05$, 
this effect would shift the \qvtx\ offset by $0.0006$ of the RMS width
of the \qvtx\ distribution. This shift is much smaller than the
statistical error on the vertex charge offset in the data, and does not
present an important additional source of systematic error.

The production flavour 
mis-tag probabilities as a function of \qt\ and \qm\ were also tested in
the data, and found to be correct to precisions of 2.6\,\% and 10\,\%,
respectively (see Section~\ref{s:bprod}).
These uncertainties were translated into errors on
\acp\ and \ccp\ by scaling the values of \qt\ and \qm\ by $\pm 2.6\,\%$ and 
$\pm 10\,\%$ as in \cite{opaljpks} and repeating the fits. 

\item[Reconstruction asymmetry:] The fit assumes that the secondary vertex
reconstruction efficiencies in the M-tag hemisphere are equal for 
\bzero\ and \bzerobar\ mesons. The track reconstruction asymmetries
mentioned in section~\ref{s:bprod} may potentially introduce an 
efficiency asymmetry,
since the vertex reconstruction requires a secondary vertex with at least
three tracks, and the sign of the highest momentum tracks will be 
different for \bzero\ and \bzerobar\ mesons. However, the track
reconstruction asymmetries in the Monte Carlo, which are somewhat
larger than those in the data, lead to no significant efficiency
asymmetry for reconstructing secondary vertices.

\item[b fragmentation:] The proper time reconstruction depends slightly on
the average energy of the b hadrons. This effect was assessed by
reweighting Monte Carlo events and repeating the fit, as discussed in
Section~\ref{ss:lfsyst}.

\item[Time resolution:] The fit is rather insensitive to the reconstructed 
proper time resolution, since the effects of CP violation are characterised
by a time scale $t\approx \pi/\dmd$ which is much larger than the average 
proper time resolution. The sensitivity was assessed by varying the
width of each Gaussian in the resolution function by $\pm 10\,\%$, varying the
fraction of the wider Gaussian by $\pm 50\,\%$ and by using an alternative
resolution function parameterisation derived from a Monte Carlo sample
with $10\,\%$ degraded tracking resolution, as in Section~\ref{ss:lfsyst}. 
\end{description}

The fit was tested on Monte Carlo by introducing non-zero values
of \acp\ and \ccp, and checking that the fit correctly reproduced the
input asymmetries. The fit errors were also verified to be correct by
splitting the Monte Carlo input into several sub-samples and studying
the distributions of fitted outputs.

Including all systematic errors, the value of \acp\ was determined from the
one parameter fit to be $\acp=\acpval\pm\acpstat\pm\acpsyst$. The 
value of \ccp\ was determined from the two parameter fit to be
$\ccp=\ccpval\pm\ccpstat\pm\ccpsyst$. The result for \acp\ can be translated
into a measurement of \repsb\ using equation~\ref{e:repsb}, and gives
\[
\repsb = \epsbval \pm \epsbstat \pm \epsbsyst
\]
where the first error is statistical and the second systematic.

\subsection{Fit and results for $\rm\bf (\Delta\tau/\tau)_b$}

The same data sample was used to measure the fractional 
difference between b and \bqbar\ hadron lifetimes. This was done by 
dividing the data into 20 bins of b/\bqbar\ hadron purity using the 
tagging variable \qe\ and simultaneously 
fitting all the reconstructed proper time distributions.
The expected proper time distribution $F_j(t)$ in bin $j$ of the tagging
variable \qe\ $(-1<\qe<1)$ is given by:
\[
F_j(t)=\int_0^\infty P_j(t') R(t-t') dt'
\]
where $P(t')$ describes the true proper time distribution of the events
and $R(t-t')$ is the time resolution function. The true proper time
distribution is given by:
\[ 
P_j(t')=(1-\fbg)
\left\{ \frac{(1+\mean{\qe}_j)}{2}\frac{1}{\taub} e^{-t'/\taub}+
\frac{(1-\mean{\qe}_j)}{2}\frac{1}{\taubbar} e^{-t'/\taubbar} \right\}
+ \fbg \left\{ \fdel\,\delta(t')+(1-\fdel)\frac{1}{\taubg}e^{-t'/\taubg}
\right\}
\]
where \fbg\ is the fraction of non-\bbbar\ background in the data
sample and $\mean{\qe}_j$ is the average value of \qe\ in the tagging bin
$j$. The parameters \taub\ and \taubbar\ are the lifetimes of b and \bqbar\
hadrons, related to the  average lifetime \tauav\ and fractional difference
\deltab\ by equation~\ref{e:deltab}. The non-\bbbar\ background was
modelled by two components---a fraction \fdel\ at $t'=0$ and the
remainder having a lifetime \taubg, both distributions being convolved
with the same resolution function as the signal. 

A two parameter $\chi^2$ fit was performed to the 20 reconstructed
proper time distributions in 0.5\,ps bins between 1 and 15\,ps,
fitting the values of \deltab\ and \tauav. 
The region below 1\,ps was excluded from the fit because the function 
$F_j(t)$ does not give a good representation of the data in this
region. This is because of the effects of the primary vertex on the 
resolution function $R(t-t')$ discussed in
Section~\ref{ss:cpvtx}. Such effects have to be taken into account
when fitting the lifetime itself \cite{opalbinc},
 but are not important when searching
for a difference in b and \bqbar\ hadron lifetimes.

In the fit, the background fraction was set to $\fbg=13.4\pm 3.0$\,\%
as in Section~\ref{ss:cpres}. The background parameters were taken
from the Monte Carlo and set to $\fdel=0.87$ and $\taubg=2.9\,$ps.

The results of the fit were
\[
\deltabbg = \delbval \pm \delbstat \\
\]
and $\tauav=\tauavval \pm \tauavstat\rm\,ps$,
where the errors are statistical only. The result for \tauav\ has
large systematic errors of the order of 0.1\,ps associated with the 
resolution function and
should not be interpreted as a measurement of the average b hadron lifetime.

The distributions of proper time in four different ranges of \qe,
together with the fit results and expectation for $\deltab=0.2$, are shown in
Figure~\ref{f:deltat}. The fit describes the data well, and 
no evidence for a difference between b and \bqbar\ hadron lifetimes is
seen.

\epostfig{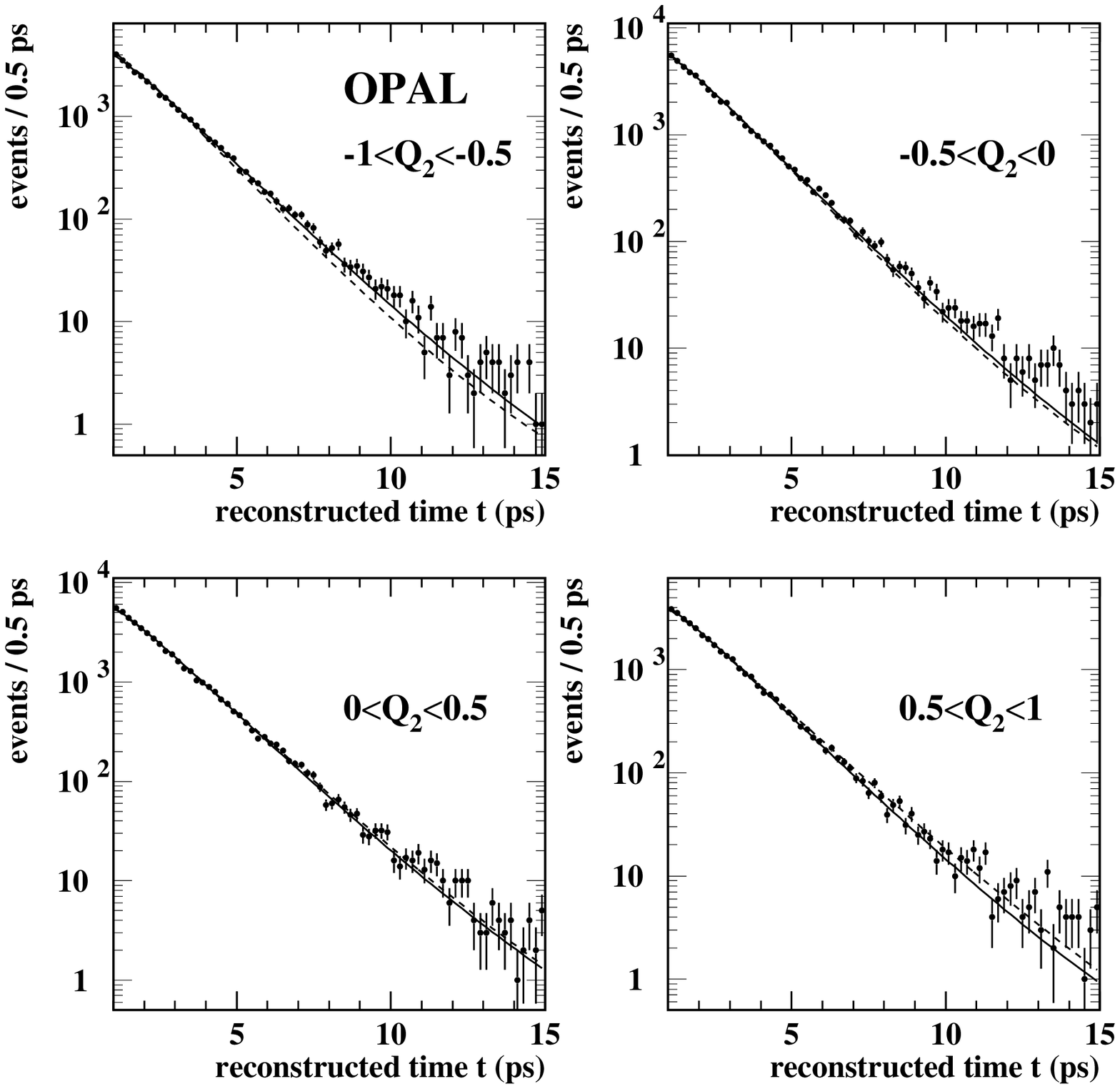}{f:deltat}{Distributions of reconstructed proper
  time $t$ ($1<t<15$\,ps) 
  in four bins of tag variable \qe. The data are shown by the
  points with error bars, and the prediction of the fit by the solid
  lines. The expected distributions for a 20\,\% difference between b and
  \bqbar\ hadron lifetimes ($\deltab=0.2$) are shown by the dotted lines.}

The systematic errors on \deltab\ are small, as most uncertainties
affect both lifetimes in the same way. They are summarised in 
Table~\ref{t:delsyst}. Most of them are similar to those of the
CP-violation analysis described in Section~\ref{ss:cpres}, and were
evaluated in the same way. The uncertainty from the background lifetime
was evaluated by varying  the fraction of background at $t'=0$ from 82\,\%
to 92\,\%, and by varying the lifetime of the other component from
2.4\,ps to 3.4\,ps. No significant effect was seen.

\begin{table}
\centering

\begin{tabular}{l|c}\hline\hline
Source & $\Delta(\deltab)$ \\ \hline
Background fraction & 0.000 \\
Background lifetime & 0.000 \\
Flavour tagging offsets & 0.001 \\
Flavour tagging mis-tag & 0.008 \\
b fragmentation & 0.000 \\
Time resolution & 0.001 \\ \hline
Total & \delbsyst \\
\hline
\end{tabular}
\caption{\label{t:delsyst} Systematic errors on the measured value of
  \deltab.}
\end{table}

The total systematic error on \deltab\ is $\pm\delbsyst$, and is dominated by
the uncertainty on the flavour mis-tag rates.
Additionally, it was checked that the fit was stable with respect to
variations of the minimum and maximum time cuts. The fit was tested
on Monte Carlo by reweighting so as to introduce variations between
the b and \bqbar\ hadron lifetimes, and checking that the fit
recovered the correct values of \deltab. The errors were checked by
splitting the Monte Carlo into subsamples, as before.

\newpage

\section{Summary and conclusions}\label{s:conc}

The lifetimes of the \bplus\ and \bzero\ mesons, and their ratio, 
have been measured
using a technique based on reconstructed secondary vertices. From a
sample of \nhadthree\ hadronic \zb\ decays collected by the OPAL detector
between 1993 and 1995, the results
\begin{eqnarray*}
\taubp & = & \tpval \pm \tpstat \pm \tpsyst\rm\,ps \\
\taubz & = & \tzval \pm \tzstat \pm \tzsyst\rm\,ps \\
\taubr & = & \trval \pm \trstat \pm \trsyst 
\end{eqnarray*}
were obtained, where in each case the first error is statistical and 
the second systematic. These results are in agreement with other
measurements from LEP, SLD and CDF
\cite{opalsemil,bsemil,delphitop,bvtx}, and with the world average
values of $\taubp=1.65\pm 0.04\rm\,ps$ and  $\taubz=1.56\pm 0.04\rm\,ps$
\cite{pdg98}. The result for the \bplus\ lifetime is the most precise 
determination to date.

Using a similar technique, an inclusive sample of b hadron decays has
been used to search for CP and CPT violation effects. No such effects
have been seen. From the time dependent asymmetry of inclusive \bzero\
decays, the CP violation parameter has been determined to be
\[
\repsb = \epsbval \pm \epsbstat \pm \epsbsyst \ .
\]
This result agrees with the OPAL
measurement using semileptonic b decays: 
$\repsb=0.002\pm 0.007\pm 0.003$ \cite{opaldms}, and is also in
agreement with other less precise results from CLEO and CDF
\cite{oldeb}.

The fractional difference in the lifetimes of b and \bqbar\ 
hadrons has also been measured to be
\[
\deltabbg \equiv \frac{\rm\tau (b\ hadron) - \tau (\bqbar\ hadron)}
{\rm\tau (average)} = \delbval \pm \delbstat \pm \delbsyst \ .
\]
This is the first published analysis to test the equality of the 
b and \bqbar\ hadron lifetimes.

\section*{Acknowledgements}

We thank D.~Silvermyr for his contribution to this analysis. \\
We particularly wish to thank the SL Division for the efficient operation
of the LEP accelerator at all energies
 and for their continuing close cooperation with
our experimental group.  We thank our colleagues from CEA, DAPNIA/SPP,
CE-Saclay for their efforts over the years on the time-of-flight and trigger
systems which we continue to use.  In addition to the support staff at our own
institutions we are pleased to acknowledge the  \\
Department of Energy, USA, \\
National Science Foundation, USA, \\
Particle Physics and Astronomy Research Council, UK, \\
Natural Sciences and Engineering Research Council, Canada, \\
Israel Science Foundation, administered by the Israel
Academy of Science and Humanities, \\
Minerva Gesellschaft, \\
Benoziyo Center for High Energy Physics,\\
Japanese Ministry of Education, Science and Culture (the
Monbusho) and a grant under the Monbusho International
Science Research Program,\\
Japanese Society for the Promotion of Science (JSPS),\\
German Israeli Bi-national Science Foundation (GIF), \\
Bundesministerium f\"ur Bildung, Wissenschaft,
Forschung und Technologie, Germany, \\
National Research Council of Canada, \\
Research Corporation, USA,\\
Hungarian Foundation for Scientific Research, OTKA T-016660, 
T023793 and OTKA F-023259.\\


\begin{thebibliography}{99}

\bibitem{pdg98}
Particle Data Group, C.~Caso~\etal, \EPJ{3}{1998}{1}.

\bibitem{spec}
For example: \\
G.~Alterelli and S.~Petrarca, \PLB{261}{1991}{303};\\
I.~Bigi, \PLB{169}{1986}{191};\\
I.~Bigi and N.G.~Uraltsev, \PLB{280}{1992}{271};\\
M.~Neubert and C.T.~Sachrajda, \NPB{483}{1997}{339}.

\bibitem{opalsemil}
OPAL collaboration, R.~Akers~\etal, \ZPC{67}{1995}{379}.

\bibitem{bsemil}
ALEPH collaboration, D.~Buskulic~\etal, \ZPC{71}{1996}{31}; \\
CDF collaboration, F.~Abe~\etal, \PRL{76}{1996}{4462}; \\
CDF collaboration, F.~Abe~\etal, \PRD{58}{1998}{092002}; \\
DELPHI collaboration, P.~Abreu~\etal, \ZPC{68}{1995}{13}; \\
DELPHI collaboration, P.~Abreu~\etal, \ZPC{74}{1997}{19}.

\bibitem{delphitop}
DELPHI collaboration, W.~Adam~\etal, \ZPC{68}{1995}{363}.

\bibitem{bvtx}
SLD collaboration, K.~Abe~\etal, \PRL{79}{1997}{590};\\
L3 collaboration, M.~Acciarri~\etal, 
`Upper Limit on the Lifetime Difference of Short- and
Long-Lived \bs\ Mesons', CERN-EP/98-127, accepted by Phys.\ Lett.\ B.

\bibitem{rebimb} 
V.A.~Kostelecky and R.~Potting, \PRD{51}{1995}{3923}.

\bibitem{rvk}
V.A.~Kostelecky and R.~Van Kooten, \PRD{54}{1996}{5585}.

\bibitem{oldeb}
CLEO collaboration, J.~Bartelt~\etal, \PRL{71}{1993}{1680};\\
CDF collaboration, F.~Abe~\etal, \PRD{55}{1997}{2546}.

\bibitem{opaldms}
OPAL collaboration, K.~Ackerstaff \etal, \ZPC{76}{1997}{401}.

\bibitem{rjh}
For example: R.~Hawkings, `CP violation in B decays at LEP', talk
presented at the 3rd International Conference on Hyperons, Charm and 
Beauty Hadrons, 2nd July 1998, Genova, Italy, to be published in 
Nucl.\ Phys.\ B.

\bibitem{cpinc}
A.~Acuto and D.~Cocolicchio, \PRD{47}{1993}{3945};\\
M.~Beneke, G.~Buchalla and I.~Dunietz, \PLB{393}{1997}{132}.

\bibitem{na14}
NA\,14/2 collaboration, M.P.~Alvarez~\etal, \ZPC{47}{1990}{539}.

\bibitem{opaldet}
OPAL collaboration, K.~Ahmet~\etal, \NIM{A305}{1991}{275}.

\bibitem{opalsi2d}
P.P.~Allport~\etal, \NIM{A324}{1993}{34}.

\bibitem{opalsi3d}
P.P.~Allport~\etal, \NIM{A346}{1994}{476}.

\bibitem{evtsel}
OPAL collaboration, G.~Alexander~\etal, \ZPC{52}{1991}{175}.

\bibitem{jetcone}
OPAL collaboration, R.~Akers~\etal, \ZPC{63}{1994}{197}.

\bibitem{jetset}
T.~Sj\"ostrand, \CPC{82}{1994}{74}.

\bibitem{jetsetopt}
OPAL collaboration, G.~Alexander~\etal, \ZPC{69}{1996}{543}.

\bibitem{fpeter}
C.~Peterson, D.~Schlatter, I.~Schmitt and P.~Zerwas, \PRD{27}{1983}{105}.

\bibitem{gopal}
J.~Allison~\etal, \NIM{A317}{1992}{47}.

\bibitem{opalrb}
OPAL collaboration, G.~Abbiendi \etal, `A Measurement of \rb\ using 
a Double Tagging Method', CERN-EP/98-137, accepted by Eur.\ Phys.\
J.\ C.

\bibitem{elecid}
OPAL collaboration, G.~Alexander~\etal, \ZPC{70}{1996}{357}.

\bibitem{muonid}
OPAL collaboration, P.D.~Acton~\etal, \ZPC{58}{1993}{523}.

\bibitem{opaldil}
OPAL collaboration, R.~Akers~\etal, \ZPC{66}{1995}{555}.

\bibitem{opaljpks}
OPAL collaboration, K.~Ackerstaff~\etal, \EPJ{5}{1998}{379}.

\bibitem{opalbstar}
OPAL collaboration, R.~Akers~\etal, \ZPC{66}{1995}{19}.

\bibitem{opaldstar}
OPAL collaboration, R.~Akers~\etal, \PLB{327}{1994}{411}.

\bibitem{beamspot}
OPAL collaboration, P.D.~Acton~\etal , \ZPC{59}{1993}{183};\\
OPAL collaboration, R.~Akers~\etal , \PLB{338}{1994}{497}.

\bibitem{opalbinc} 
OPAL collaboration, K.~Ackerstaff~\etal, \ZPC{73}{1997}{397}. 

\bibitem{hfew}
The LEP experiments, ALEPH, DELPHI, L3 and OPAL,
\NIM{A378}{1996}{101}.\\
Updated averages are described in `Presentation of LEP Electroweak
Heavy Flavour Results for Summer 1998 Conferences', LEPHF 98-01
(see {\tt http://www.cern.ch/LEPEWWG/heavy/} ).

\bibitem{dun2}
I.~Dunietz, `CP Asymmetries in Semi-inclusive \bzero\ Decays', 30th
June 1998, FERMILAB-PUB-97/323-T, hep-ph/9806521.

\bibitem{superw}
D.~Cocolicchio and L.~Maiani, \PLB{291}{1992}{155};\\
J.~Gerard and T.~Nakada, \PLB{261}{1991}{474};\\
J.~Liu and L.~Wolfenstein, \PLB{197}{1987}{536}.

\end{thebibliography}
\end{document}